\documentclass[twocolumn]{svjour3}          
\smartqed  
\usepackage{graphicx}
\usepackage{amssymb}
\usepackage{amsmath}
      \newcommand{\etal}{et~al. }
 \journalname{}
\begin{document}

\title{Modelling and simulation of adhesive curing processes in bonded piezo metal composites}


\author{Ralf Landgraf \and
            Robert Scherzer \and
            Martin Rudolph \and
            J\"orn Ihlemann
}

\institute{R. Landgraf \and
             R. Scherzer \and
             M. Rudolph \and
             J. Ihlemann  \at
              Department of Solid Mechanics, \\
              Institute of Mechanics and Thermodynamics, \\
              Faculty of Mechanical Engineering, \\
              Chemnitz University of Technology, 
              Chemnitz, Germany \\[1mm]
              R. Landgraf \at 
              \email{ralf.landgraf@mb.tu-chemnitz.de} 
              \\[1mm]
              R. Scherzer \at \email{robert.scherzer@mb.tu-chemnitz.de}
              \\[1mm]
              M. Rudolph \at \email{martin.rudolph@mb.tu-chemnitz.de}
              \\[1mm]
              J. Ihlemann \at \email{joern.ihlemann@mb.tu-chemnitz.de}
}

\maketitle

\begin{abstract}
This work deals with the modelling and simulation of curing phenomena in adhesively bonded piezo metal composites which consist of an adhesive layer with integrated piezoelectric module and two surrounding metal sheet layers. In a first step, a general modelling framework is proposed which is able to represent curing phenomena in polymers at finite strains. Based on this formulation, a concretized model is deduced for the simulation of curing in one specific epoxy based adhesive. Here, appropriate material functions are specified and the thermodynamic consistency is proved. Regarding the finite element implementation, a numerical integration scheme and a new approach for the consideration of different initial conditions are provided. Finally, finite element simulations of a newly proposed manufacturing process for the production of bonded piezo metal composite structures are conducted. A deep drawing process of the composite with uncured adhesive layer and the subsequent adhesive curing are investigated. 

\end{abstract}

\newbox\JIGAMMa
\newbox\JIGAMMb
\newbox\JIGAMMc
\newbox\JIGAMMd
\newbox\JIGAMMz
\newbox\LLbox
\newbox\LLboxh
\newbox\SLhilfbox
\newbox\SLubox
\newbox\SLobox
\newbox\SLergebnis
\newbox\TENbox
\newif\ifSLoben
\newif\ifSLunten
\newdimen\JIGAMMdimen
\newdimen\JIhsize\relax\JIhsize=\hsize
\newdimen\SLrandausgleich
\newdimen\SLhoehe
\newdimen\SLeffbreite
\newdimen\SLuvorschub
\newdimen\SLmvorschub
\newdimen\SLovorschub
\newdimen\SLsp
\def\rhotilzurho{{\JI\frac{\STAPEL\varrho!^\SLtilde\!}{\JI\!\varrho}}}
\def\rhozurhotil{{\JI\varrho\over\JI\STAPEL\varrho!^\SLtilde}}
\def\pkt{\cdot}
\def\ppkt{\mathbin{\mathord{\cdot}\mathord{\cdot}}}
\setbox\JIGAMMa = \hbox{$\scriptscriptstyle c$} \setbox\JIGAMMz =
\hbox{\hskip-.35pt\vrule width .25pt\hskip-.35pt
                        \vbox to1.2\ht\JIGAMMa{\vskip-.125pt
                             \hrule width1.2\ht\JIGAMMa height.25pt
                             \vfill
                             \hrule width1.2\ht\JIGAMMa height.25pt
                             \vskip-.125pt}%
                        \hskip-.125pt\vrule width .25pt\hskip-.125pt}
\def\Oldroy#1#2#3{\STAPEL{#1}!_\SLstrich!_\SLstrich!^\circ{}
                  \ifx #2,{}_{\copy\JIGAMMz}%
                  \else \mskip1mu{}^{\copy\JIGAMMz}\fi
                  \mskip1mu\ifx #3,{}_{\copy\JIGAMMz}%
                           \else {}^{\copy\JIGAMMz}\fi }
\def\OP#1#2{\ifnum#1=1{\rm S}
            \else\ifnum#1=2{\rm S}^\JIv
                 \else\ifnum#1=3{\rm S}^T
                      \else{{\rm S}^T}^\JIv
            \fi\fi\fi\LL{{#2}}\RR}
\edef\JIminus{{\setbox\JIGAMMa=\hbox{$\scriptstyle x$}%
           \hbox{\hskip .10\wd\JIGAMMa
                 \vbox{\hrule width .6\wd\JIGAMMa height .07\wd\JIGAMMa
                       \vskip.53\ht\JIGAMMa}%
                 \hskip .10\wd\JIGAMMa}}}
\edef\JIv{{\JIminus 1}}
\def\JI{\displaystyle}
\def\JIha{{1\over 2}}
\def\JIfolgt{\quad\Rightarrow\quad}
\def\LL#1\RR{\setbox\LLbox =\hbox{\mathsurround=0pt$\displaystyle
                                              \left(#1\right)$}%
       \setbox\LLboxh=\hbox{\mathsurround=0pt%
                  $\displaystyle{\left(%
                      \vrule width 0pt height\ht\LLbox depth\dp\LLbox
                      \right)}$}%
       \left(\hskip-.3\wd\LLboxh\relax\copy\LLbox
              \hskip-.3\wd\LLboxh\relax\right)}
\def\ZBOX#1#2#3{\def#3{}%
                \setbox#1 = #2
                \def#3{ to \wd#1}%
                \setbox#1 = #2}
\def\SLdreieck{\setbox\TENbox=\hbox{\fontscsy\char 52}
                 \dp\TENbox = 0pt
                 \hbox{\hskip -2\SLrandausgleich
                       \box\TENbox
                       \hskip -2\SLrandausgleich}}
\def\SLtilde{\setbox\TENbox=\hbox{\fontscex\char 101}
                    \vbox{\vskip-.03\ht\TENbox
                          \hbox{\hskip -1\SLrandausgleich
                                \copy\TENbox
                                \hskip -1\SLrandausgleich}
                          \vskip -.86\ht\TENbox}}
\def\SLstrich{\vrule width \SLeffbreite height.4pt}
\def\SLpunkt{{\vbox{\hbox{$\displaystyle.$}\vskip.03cm}}}
\def\SLabstand{\vskip .404pt}
\def\SLzwischen{\vskip 1.372pt}
\font\fontscsy=cmsy6 \font\fontscex=cmex10 scaled 1200
\def\STAPEL#1{\def\SLkern{#1}%
              \futurelet\next\SLpruef
               A_0   _0    :B_0   _-.17 :C_.05 _-.15 :D_0   _-.2
              :E_0   _-.2  :F_0   _-.21 :G_0   _-.15 :H_0   _-.23
              :I_.2  _.15  :J_.05 _-.1  :K_0   _-.22 :L_0   _-.1
              :M_0   _-.23 :N_0   _-.25 :O_.05 _-.2  :P_0   _-.21
              :Q_.05 _-.2  :R_0   _-.03 :S_0   _-.15 :T_.2  _0
              :U_.1  _-.1  :V_.1  _-.15 :W_.1  _-.2  :X_0   _-.22
              :Y_.16 _-.15 :Z_0   _-.25
              :a_.05 _-.05 :b_.05 _0    :c_.05 _.05  :d_0   _-.05
              :e_.07 _0    :f_0   _-.15 :g_.04 _-.2  :h_0   _-.07
              :i_.05 _0    :j_.08 _-.1  :k_0   _-.1  :l_.2  _.15
              :m_0   _-.1  :n_0   _-.1  :o_0   _-.1  :p_.15 _0
              :q_.1  _0    :r_.1  _-.1  :s_0   _-.2  :t_.1  _.05
              :u_0   _-.1  :v_0   _-.2  :w_0   _-.2  :x_.04 _-.14
              :y_.15 _-.05 :z_0   _-.15
              :\mit\Phi_.08 _-.1    :\mit\Omega_0 _-.2   :\varXi_.00 _-.2
              :\alpha_0 _-.2        :\gamma_.1 _-.1      :\varepsilon_.1 _-.1
              :\epsilon_.05 _-.05   :\eta_.05 _-.15      :\lambda_0 _0
              :\mu_0 _-.25          :\nu_0 _-.2          :\varSigma_-.03 _-.2
              :\varrho_.00 _-.2     :\sigma_.1 _-.2      :\tau_.15 _-.15
              :\theta_.2 _0          :\nabla_.1 _-.2
              :\varphi_.2 _-.1      :\omega_.1 _-.1      :\mit\Gamma_-.1 _-.1
              :\Lambda_0 _0         :\Gam_0 _0           :\Lam_0 _0
              :{\cal K}_1 _0
              :\SLsuchende
              \def\SLtrick{\noexpand\SLtrick\noexpand}%
                \def\SLdummy{\noexpand\SLdummy}%
                \edef\SLoboxinhalt{}\edef\SLuboxinhalt{}%
                \SLobenfalse\SLuntenfalse
                \futurelet\next\SLsuchruf}
  \def\SLsuchruf{\ifx\next !\let\next\SLexpand
                 \else\let\next\SLerzeug\fi\next}
  \def\SLexpand#1#2#3{\ifx #2\sb\ifSLunten\let\SLspeicher\SLuboxinhalt
                   \else\def\SLspeicher{\SLtrick\SLabstand}\fi
                   \edef\SLuboxinhalt{%
                       \SLspeicher
                       \SLtrick\SLzwischen
                       \hbox\SLdummy{\hfil\mathsurround=0pt
$\SLtrick\scriptstyle\SLtrick#3$%
                                     \hfil}}%
                   \SLuntentrue%
                      \else\ifSLoben\let\SLspeicher\SLoboxinhalt
                   \else\def\SLspeicher{\SLtrick\SLabstand}\fi
                   \edef\SLoboxinhalt{%
                       \hbox\SLdummy{\hfil\mathsurround=0pt
$\SLtrick\scriptstyle\SLtrick#3$%
                                     \hfil}%
                       \SLtrick\SLzwischen
                       \SLspeicher}%
                   \SLobentrue\fi\futurelet\next\SLsuchruf}
  \def\SLerzeug{\def\SLtrick{}
                \setbox\SLhilfbox=\hbox{$\displaystyle{E}$}%
                \SLrandausgleich=.04\wd\SLhilfbox
                      \setbox\SLhilfbox=%
                         \hbox{\hskip -1\SLrandausgleich
                          \mathsurround=0pt$\displaystyle{\SLkern}$%
                               \hskip -1\SLrandausgleich}%
                      \SLhoehe = \ht\SLhilfbox
                      \advance\SLhoehe by \dp\SLhilfbox
                      \SLeffbreite = \wd\SLhilfbox
                      \advance\SLeffbreite by \SLab\SLhoehe
                      \ZBOX\SLubox{\vbox{\offinterlineskip
                                         \SLuboxinhalt
                                         \hrule height 0pt}}\SLdummy
                      \ZBOX\SLobox{\vbox{\offinterlineskip
                                         \SLoboxinhalt
                                         \hrule height 0pt}}\SLdummy
                      \SLsp = \SLzu\SLhoehe
                      \advance\SLsp by -.5\SLeffbreite
                      \SLuvorschub = -1\SLsp
                      \advance\SLuvorschub by -.5\wd\SLubox
                      \SLovorschub = -1\SLsp
                      \advance\SLovorschub by -.5\wd\SLobox
                      \advance\SLovorschub by .26\SLhoehe
                      \ifdim\SLuvorschub > \SLovorschub
                         \SLsp = \SLovorschub
                      \else
                         \SLsp = \SLuvorschub
                      \fi
                      \ifdim\SLsp < 0pt%
                         \advance\SLuvorschub by -1\SLsp
                         \SLmvorschub = -1\SLsp
                         \advance\SLovorschub by -1\SLsp
                      \else
                         \SLmvorschub = 0pt
                      \fi
                      \setbox\SLergebnis = \hbox{%
                         \offinterlineskip
                         \hskip\SLrandausgleich\relax
                         \vbox to 0pt{%
                            \vskip -1\ht\SLobox
                            \vskip -1\ht\SLhilfbox
\hbox{\hskip\SLovorschub\copy\SLobox\hfil}%
                            \hbox{\hskip\SLmvorschub\copy\SLhilfbox
                                  \hfil}%
\hbox{\hskip\SLuvorschub\copy\SLubox\hfil}%
                            \vss}%
                         \hskip\SLrandausgleich}%
                      \SLsp = \dp\SLhilfbox
                      \advance\SLsp by \ht\SLubox
                      \dp\SLergebnis = \SLsp
                      \SLsp = \ht\SLhilfbox
                      \advance\SLsp by \ht\SLobox
                      \ht\SLergebnis = \SLsp
                      \box\SLergebnis{}}
  \def\SLpruef{\ifx\next\SLsuchende\def\SLzu{0}\def\SLab{0}%
                  \def\next##1\SLsuchende{\relax}%
               \else\let\next\SLvergl
               \fi\next}
  \def\SLvergl#1_#2_#3:{\def\SLv{#1}%
                        \ifx\SLkern\SLv\def\SLzu{#2}\def\SLab{#3}%
\def\next##1\SLsuchende{\relax}%
                        \else\def\next{\futurelet\next\SLpruef}
                        \fi\next}
\def\PKT#1{#1!^\SLpunkt}

\newbox\minusbox
\def\minus{\mathchoice{\minusarb\displaystyle}%
                      {\minusarb\textstyle}%
                      {\minusarb\scriptstyle}%
                      {\minusarb\scriptscriptstyle}}
  \def\minusarb#1{\setbox\minusbox=\hbox{$#1x$}%
                  \hbox{\hskip .10\wd\minusbox
                        \vbox{\hrule width .6\wd\minusbox
                                     height .07\wd\minusbox
                              \vskip.53\ht\minusbox}%
                        \hskip .10\wd\minusbox}}

\def\Basis{\STAPEL e!_\SLstrich}
\def\C{\STAPEL C!_\SLstrich!_\SLstrich}
\def\X{\STAPEL X!_\SLstrich!_\SLstrich}
\def\Cinv{\STAPEL C!_\SLstrich!_\SLstrich^{-1}}
\def\Cg{\STAPEL C!_\SLstrich!_\SLstrich!^\SLstrich}
\def\CD{\STAPEL C!_\SLstrich!_\SLstrich!^\SLdreieck}
\def\XD{\STAPEL X!_\SLstrich!_\SLstrich!^\SLdreieck}
\def\ECKMT{\STAPEL M!_\SLstrich!_\SLstrich!^\SLdreieck^T}
\def\CgD{\STAPEL C!_\SLstrich!_\SLstrich!^\SLstrich!^\SLdreieck}
\def\BD{{\STAPEL C!_\SLstrich!_\SLstrich!^\SLdreieck}{^{\JIv}}}
\def\BDn#1{{\STAPEL C!_\SLstrich!_\SLstrich!^\SLdreieck}{^{\JIv}_{#1}}}
\def\Ttil{\STAPEL T!_\SLstrich!_\SLstrich!^\SLtilde}
\def\rhotil{\JI\STAPEL\varrho!^\SLtilde\!}
\def\rhotilzurho{{\JI\frac{\STAPEL\varrho!^\SLtilde\!}{\JI\!\varrho}}}
\def\rhozurhotil{{\JI\varrho\over\JI\STAPEL\varrho!^\SLtilde}}
\def\Skal#1{\STAPEL {#1}}
\def\Vek#1{\STAPEL {#1}!_\SLstrich}
\def\Ten2#1{\STAPEL {#1}!_\SLstrich!_\SLstrich}
\def\F{\STAPEL F!_\SLstrich!_\SLstrich}
\def\Fg{\STAPEL F!_\SLstrich!_\SLstrich!^\SLstrich}
\def\e{\STAPEL e!_\SLstrich!_\SLstrich}
\def\b{\STAPEL b!_\SLstrich!_\SLstrich}
\def\D{\STAPEL D!_\SLstrich!_\SLstrich}
\def\I{\STAPEL I!_\SLstrich!_\SLstrich}
\def\I{\STAPEL I!_\SLstrich!_\SLstrich}
\def\L{\STAPEL L!_\SLstrich!_\SLstrich}
\def\P{\STAPEL P!_\SLstrich!_\SLstrich}
\def\CnOldWed{\overset{\circ{}}{\Big(\STAPEL{C}!_\SLstrich!_\SLstrich_2\Big)}
                  \mskip1mu\mskip1mu
                  \mskip1mu {}^{\hat{\copy\JIGAMMz}}
                  \mskip1mu {}_{\hat{\copy\JIGAMMz}}}
\def\CnOldWedred{\STAPEL{C}!_\SLstrich!_\SLstrich!^\circ{}_2}

\def\CD{\C!^\SLdreieck}
\def\CDn#1{\C!^\SLdreieck_#1}
\def\X{\Ten2 X}
\def\XDn#1{\X!^\SLdreieck_#1}
\def\Cauchy{\STAPEL \sigma!_\SLstrich!_\SLstrich}
\def\Cn#1{\C_{#1}}
\def\Ln#1{\L_{#1}}
\def\h{\STAPEL h!_\SLstrich!_\SLstrich}
\def\Fn#1{\F_{#1}}
\def\qtil{\STAPEL q!_\SLstrich!^\SLtilde}
\def\nabtil{\STAPEL \nabla!^\SLtilde!_{\hspace{0.5ex}\SLstrich}}
\def\omtil{\STAPEL \omega!_\SLstrich!^\SLtilde}
\def\curetil{\STAPEL q!^\SLtilde}
\def\thetatil{\STAPEL \theta!^\SLtilde}
\def\Ktil{\STAPEL {\cal K}!^\SLtilde}
\def\Vtil{\STAPEL V!^\SLtilde}
\newcommand{\IntGtil}[1]
   {\displaystyle{{\Big.^{\widetilde{\mathcal{G}}}}
      \hspace{-1.5ex}\int #1 \ {\rm d}\widetilde{V}}}
\newcommand{\IntGtilN}[2]
   {\displaystyle{{\Big.^{\widetilde{\mathcal{G}}_{#1}}}
      \hspace{-1.5ex}\int #2 \ {\rm d}\widetilde{V}_{#1}}}
\newcommand{\IntGhat}[1]
   {\displaystyle{{\Big.^{\widehat{\mathcal{G}}}}
      \hspace{-1.5ex}\int #1 \ {\rm d}\widehat{V}}}
\newcommand{\nadel}[1]
   {\big(\hspace{-3pt}\big(\hspace{1pt} #1\hspace{1pt}\big) \hspace{-3pt}\big)}
  
\newcommand{\inv}{{\minus 1}}
\newcommand{\invT}{{\minus\T}}    

\def\indLT#1#2%
  {\mathop{}%
   \mathopen{\vphantom{#2}}^{\scriptscriptstyle #1}%
   \kern-\scriptspace%
   #2}
%
\section{Introduction}
\label{sec:Introduction}

Motivated by the desire for continuous improvement, industrial countries constantly aim to develop innovative concepts and products of highest standards. Within this context, lightweight construction and smart structures are doubtless crucial keywords nowadays. Within the last decade, a number of new developments based on lightweight concepts were successfully established in nearly all fields of engineering \cite{Wiedemann_Sinapius_2013}. One very challenging aspect for the implementation of such new concepts is the joining technology. Thereby, adhesives are given an important role because they join the most diverse materials, not only locally but also as full-surface bonding \cite{messler_2004}. In the research field of smart structures, high importance is awarded to piezoceramic patches as they combine static structures with actuator and sensor functionality \cite{Prasad_Etal_2005}. In that context, the Piezoceramic Fibre Composites (PFC) were shown to be the most promising technology. More precisely, the Macro Fibre Composite (MFC) is the most sophisticated device yet invented \cite{Lloyd_2004}. 

\begin{figure}[ht]
	\centering
	\includegraphics[width=0.45\textwidth]{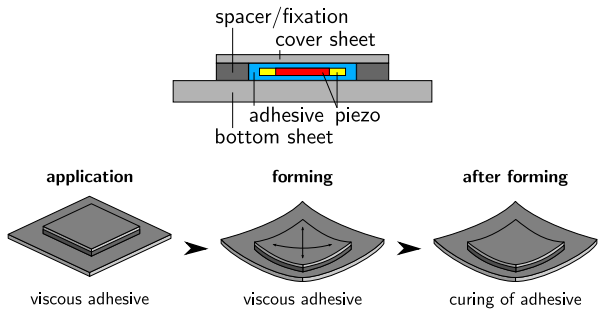}
	\caption{Schematic illustration of the piezo metal composite (top) and manufacturing process (bottom)}
	\label{fig:pic_Principle-PMC}
\end{figure}

Despite excellent properties of piezoceramic patches, the state of art is their application to the fabricated parts only after manufacturing which leads to a time and cost intense procedure \cite{Neugebauer_Etal_2010_ProdEng}. Scientific fundamentals for an economic production of active structural components are worked out in the Collaborative Research Center/Transregio "PT-PIESA". One of the pursued concepts, that is considered in this paper, is the joining of sheet metal lightweight construction and piezo elements with structural adhesives to smart Piezo Metal Composites (PMC) by an innovative ma\-nu\-fac\-tu\-ring approach. The basic idea is to merge the steps of forming and piezo application into one process such that they are no longer separated \cite{Neugebauer_Etal_2010_ProdEng,Neugebauer_Etal_2013_WGP}. A schematic representation of the approach is depicted in Fig.~\ref{fig:pic_Principle-PMC}.

Firstly, the MFC is entirely surrounded by a structural adhesive, placed inside two light metal sheets, where one of them is the sheet intended to be formed and the other is a local covering sheet, see Fig.~\ref{fig:pic_Principle-PMC} (top). A specific distance between both metal sheets is adjusted by the help of spacers. Next, the sandwich structure is formed to its final shape while the adhesive is not yet cured. During this stage, the MFC is protected from excessively high loads by a floating support. After the forming process, the adhesive cures to a solid and thereby provides a material closure between the MFC and the light metal structure in the formed state. The principal feasibility of this method could already be demonstrated in earlier studies (see, for instance, \cite{Drossel_Etal_2009_CIRP,Neugebauer_Etal_2010_ProdEng,Neugebauer_Etal_2013_WGP}). 

Within the PMC, an essential role is given to the adhesive layer. Beside the impact of the adhesive's specific material behaviour, its geometrical design (i.e. the layer thickness) is of great importance. If, on the one hand, the adhesive layer is very thin, the protective function during forming vanishes. On the other hand, if the adhesive layer is too thick, risk of overloads due to volume shrinkage processes increases. Moreover, secondary deformations of the PMC might occur and a thick adhesive layer may lead to loss of the electric field in the piezoceramic due to the additional capacity between actuator and structure \cite{Seemann_Sattel_1999}. 

First studies on the influence of the adhesive during forming have been conducted by Neugebauer \etal \cite{Neugebauer_Etal_2013_WGP}. However, curing of the adhesive has not been taken into account so far. Thus, the aim of this study is to set up a simulation tool which enables the simulation of curing phenomena in adhesives and to investigate the impact of the curing process on formed PMCs and more precisely on the embedded MFC. One essential part of this work is to provide a phenomenological model which is capable of representing the material behaviour of the adhesive during cure and in the fully cured state. The main characteristics of this model are
\begin{itemize}
  \item the description of the progress of the chemical process during cure,
  \item the modelling of dependencies of mechanical properties during the curing process as well as at different temperatures, and
  \item the prediction of volume changes which are caused by chemical shrinkage and heat expansion phenomena.
\end{itemize}
Here, different modelling approaches have been presented before (see, for example, \cite{Hossain_Etal_2009,Hossain_Etal_2010,Klinge_Etal_2012,Kolmeder_Etal_2011,Liebl_Etal_2012,Lion_Hoefer_2007,Mahnken_2013}). The basic structure of those models is similar. Beside the application to different specific materials, one further basic difference is their employed mechanical submodel. For example models of finite strain elasticity \cite{Hossain_Etal_2009},  finite strain viscoelasticity \cite{Hossain_Etal_2010,Klinge_Etal_2012,Kolmeder_Etal_2011,Lion_Hoefer_2007,Mahnken_2013} and viscoplasticity at small \cite{Liebl_Etal_2012} and finite strains \cite{Landgraf_Ihlemann_2011} have been used. In this paper, a general modelling approach which includes the main characteristics of the Lion and H\"ofer model \cite{Lion_Hoefer_2007} is presented (see Section \ref{sec:Curing_model}). However, it is formulated in a more general way. Especially, different mechanical submodels can be incorporated to represent the mechanical behaviour during curing. 

In Section~\ref{sec:Curing_model_constitutive_functions}, a particular model is introduced  which is able to capture curing phenomena of one specific two component epoxy based adhesive. To this end, appropriate constitutive material functions are chosen and the thermodynamic consistency is evaluated. Within this specification, the mechanical behaviour is represented by a combination of models of finite strain pseudo-elasticity and viscoelasticity. Furthermore, changes in volume  due to heat expansion and chemical shrinkage processes are taken into account.

The second part of this paper deals with different aspects of the finite element implementation (see Section~\ref{sec:FEM_implementation}). The numerical integration of constitutive equations as well as the derivation of appropriate stress and material tangent measures for the implementation into the finite element software \textit{ANSYS}$^{\rm TM}$ are described. Moreover, a new algorithm is presented, which addresses numerical difficulties that arise due to thermal and chemically related volume changes. The constitutive functions for the representation of heat expansion and chemical shrinkage processes are introduced with respect to specific reference values for the temperature and a degree of cure, which is an internal variable representing the progress of the curing process. If initial values for both variables differ from previously defined reference values, an immediate volume change would be computed which may lead to instant mesh distortion. The new algorithm calculates a correction and thus keeps the initial volume constant for arbitrary initial values.

Finally, the material model is applied to the simulation of curing processes in bonded PMCs which is described in Section \ref{sec:Finite_element_simulation}. Here, a finite element model of a deep drawn cup geometry is employed in a simplified manner such that only the part directly surrounding the MFC is modelled. To obtain a realistic forming simulation, the geometry of the final formed model relies on data which has been extracted from comprehensive simulations presented by Neugebauer \etal \cite{Neugebauer_Etal_2013_WGP}. This simplified approach allows for reduction of computational efforts related to complicated forming simulations and makes it possible to concentrate on phenomena which accompany the curing of the adhesive. An analysis of the strains in the MFC will highlight the benefits of the new process chain of manufacturing described above.

%
\section{Constitutive modelling of curing phenomena in polymers}
\label{sec:Curing_model}

For the mathematical representation of the phenomenological model presented in this paper, a coordinate free tensor formalism according to Ihlemann \cite{Ihlemann_2006} is used. Thereby, the rank of a tensor is denoted by the number of its underlines. To exemplify, $\Ten2 X$ and $\STAPEL K!_\SLstrich!_\SLstrich!_\SLstrich!_\SLstrich$ are second- and fourth-rank tensors, respectively.  Furthermore, the following general notations are used throughout this article:
\begin{itemize}
  \item second-rank identity tensor: $\I$, \\[-2mm]
  \item first and third principle invariant: $I_1(\Ten2 X)$ , $I_3(\Ten2 X)$, \footnote{ The first principle invariant equals the trace operator of the Cartesian coordinates $X_{ab}$, thus $I_1(\Ten2 X) = {\rm trace}[X_{ab}]$. Accordingly, the third principle invariant can be derived by the determinant, thus $I_3(\Ten2 X) = {\rm det}[X_{ab}]$}\\[-2mm]
  \item deviatoric part of a tensor: $\Ten2 X' = \Ten2 X - \frac{1}{3}\,I_1(\Ten2 X) \, \I$,\\[-2mm]
  \item unimodular part of a tensor: $\Ten2 X!^\SLstrich = I_3(\Ten2 X)^{-1/3} \, \Ten2 X$,\\[-2mm]
  \item inverse and transpose of a tensor: $\Ten2 X^\inv$ and $\Ten2 X^T$,\\[-2mm]
  \item material time derivative: $\frac{\rm d}{{\rm d}t}\Ten2 X = \Ten2 X!^\SLdreieck$.
\end{itemize}

A further tensor operation is introduced as follows. Assume two arbitrary second rank tensors $\Ten2 X$ and $\Ten2 Y$ and a symmetric second rank tensor $\Ten2 Z = \Ten2 Z^T$. Based on these, a tensor operation denoted by superscript $S_{24}$ is defined by
\begin{equation}
\label{eq:S24}
  \left( \Ten2 X \otimes \Ten2 Y\right)^{S_{24}}  \ppkt \Ten2 Z
    = \dfrac{1}{2} \left( \Ten2 X \cdot \Ten2 Z \cdot \Ten2 Y + \Ten2 Y^T \cdot \Ten2 Z \cdot \Ten2 X^T \right) \, .
\end{equation} 
In the following, the kinematics and constitutive assumptions of the general modelling approach are presented. 

\subsection{Kinematics}
\label{sec:Kinematics}
\begin{figure}[ht]
	\centering
	\includegraphics[width=0.45\textwidth]{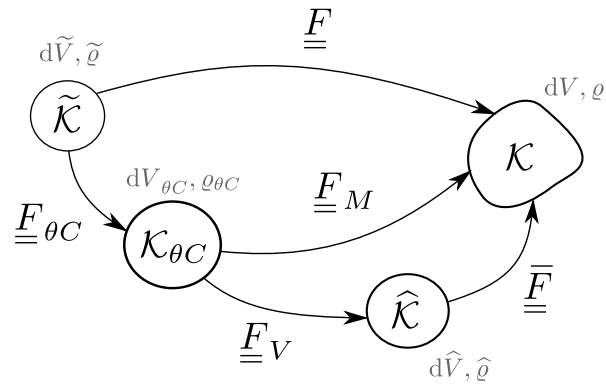}
	\caption{Multiplicative decomposition of the deformation gradient}
	\label{fig:defgrad}
\end{figure}

The phenomenological model for the representation of adhesive's curing is built up within the framework of nonlinear continuum mechanics using the deformation gradient $\F$ for the description of the underlying kinematics. The corresponding right Cauchy-Green tensor $\C$ is defined by
\begin{equation}
\label{eq:rCG}
  \C = \F^T\cdot\F \ .
\end{equation}
Furthermore, the total volume ratio is abbreviated by $J = I_3(\F) =   {\rm d} V /{\rm d}\STAPEL V!^\SLtilde $. To capture different sources of deformation, the deformation gradient gets multiplicatively decomposed as depicted in Fig.~\ref{fig:defgrad}.

Firstly, $\F$ gets decomposed into a thermochemical part $\F_{\theta C}$ and a mechanical part $\F_M$ by
\begin{equation}
\label{eq:split_mech_thermochem}
  \F = \F_{M}\cdot\F_{\theta C} \ .
\end{equation}
The thermochemical part is related to chemical shrinkage and heat expansion phenomena which are assumed to be isotropic. Thus, $\F_{\theta C}$ is an isotropic tensor 
\begin{equation}
\label{eq:FthetaC}
  \F_{\theta C} = J_{\theta C}^{\,1/3} \ \I 
  \ , \quad
  J_{\theta C} = \varphi_{\theta C}(\theta,q) \ ,
\end{equation}
where $J_{\theta C} = I_3(\F_{\theta C}) = {\rm d} V_{\theta C}/{\rm d}\STAPEL V!^\SLtilde$ is the scalar valued volume ratio which denotes the pure thermochemical volume change. This volume ratio is constituted by a function $\varphi_{\theta C}(\theta,q)$ which depends on the thermodynamic temperature $\theta$ and a variable $q$ referred to as degree of cure. A specific ansatz for  $\varphi_{\theta C}$ is provided in Section \ref{sec:Constitutive_functions_Volume}. The mechanical part of the deformation gradient $\F_M$ as well as its corresponding right Cauchy-Green tensor $\C_M$ are calculated by substituting Eq.~\eqref{eq:FthetaC}$_1$ into \eqref{eq:split_mech_thermochem} which yields
\begin{equation}
\label{eq:FM_CM}
   \F_M = J_{\theta C}^{\,-1/3} \F \ , \quad
   \C_M = \F_M^T\cdot\F_M = J_{\theta C}^{\,-2/3} \C \ .
\end{equation}
Next, the mechanical deformation gradient $\F_{M}$ is multiplicatively decomposed into $\F_V$ representing pure mechanical volume changes and a remaining isochoric (i.e. volume-preserving) part $\Fg$:
\begin{equation}
\label{eq:split_vol_isochor}
  \F_M = \Fg\cdot\F_{V} \ ,
  \quad \F_V = J_M^{\,1/3} \ \I \ .
\end{equation}
Therein, $J_M = I_3(\F_M) =  {\rm d}\widehat{V} /{\rm d}V_{\theta C}$ is the mechanical volume ratio. Substituting \eqref{eq:split_vol_isochor}$_2$ into \eqref{eq:split_vol_isochor}$_1$ yields the isochoric deformation gradient 
\begin{equation}
\label{eq:Fg}
  \Fg = J_M^{-1/3} \ \F_M = J^{\,-1/3} \ \F \ ,
\end{equation}
which exhibits the property $I_3(\Fg) = 1$. Its corresponding isochoric right Cauchy-Green tensor is calculated by
\begin{equation}
\label{eq:Cg}
   \Cg =  \Fg^T\cdot\Fg = J_M^{\,-2/3} \ \C_M= J^{-2/3} \ \C \ .
\end{equation}

At this point, all necessary aspects of the underlying kinematics have been introduced. However, for subsequent evaluations, time derivatives of different kinematic quantities have to be calculated as well. In the following, the most important relations will be summarized. 

The material time derivative of the mechanical right Cauchy-Green tensor $\C_M$ (cf. eq. \eqref{eq:FM_CM}$_2$) is given by
\begin{equation}
\label{eq:CMdot}
\begin{array}{lcl}
  \C!^\SLdreieck_M 
     =  \dfrac{\rm d}{{\rm d}t} \Big[ J_{\theta C}^{\,-2/3} \ \C\Big] \\[3mm]
  \phantom{\C!^\SLdreieck_M}
     =  J_{\theta C}^{\,-2/3} 
              \left\{ \C!^\SLdreieck 
                       - \dfrac{2}{3}  \dfrac{1}{J_{\theta C}}  
                          \left(
                              \dfrac{\partial J_{\theta C}}{\partial \theta} \dot{\theta}
                            +\dfrac{\partial J_{\theta C}}{\partial q} \dot{q}
                          \right)\C
              \right\}.
\end{array} 
\end{equation}
Moreover, the rate of the mechanical volume ratio $J_M$ equals
\begin{equation}
\label{eq:JMdot}
\begin{array}{lcl}
  \STAPEL J!^\SLpunkt_M 
    \, = \, \dfrac{\rm d}{{\rm d}t} \bigg[ \sqrt{I_3(\C_M)}\bigg] 
    \, = \, \dfrac{1}{2} \, J_M \, \C_M^\inv \ppkt \C!^\SLdreieck_M \ .
\end{array}
\end{equation}
The material time derivative of $\Cg$ (see eq. \eqref{eq:Cg}) can be expressed by
\begin{equation}
\label{eq:Cgdot}
\begin{array}{lcl}
  \Cg!^\SLdreieck 
    \, = \, \dfrac{\rm d}{{\rm d}t}\Big[J_M^{-2/3}J_{\theta C}^{-2/3} \ \C \Big]
    \, = \, \Cg\cdot \left( \C_M^\inv \cdot \C!^\SLdreieck_M\right)' \ .
\end{array}
\end{equation}

\subsection{General modelling framework}
\label{sec:Curing_model_general_framework}

In this section, a general modelling framework is introduced which defines the basic structure of the adhesives material model. To obtain a thermodynamically consist model, the second law of thermodynamics in form of the Clausius-Duhem inequality is considered. In Lagrangian representation it reads as follows
\begin{equation}
\label{eq:CDU}
  \dfrac{1}{2} \, \Ttil \ppkt \C!^\SLdreieck 
     -  \rhotil  \dot{\psi} - \rhotil \, \eta \, \dot{\theta} - \dfrac{1}{\theta} \qtil \cdot \nabtil \theta \ge 0 \ .
\end{equation}
Therein, the first term is the stress power per unit volume, $\Ttil$ is the $2^{\rm nd}$ Piola-Kirchhoff stress tensor and $\psi$ and $\eta$ are the Helmholtz free energy and the entropy, respectively, per unit mass. Furthermore, $\rhotil$ is the mass density and $\qtil$ is the heat flux vector, both defined on the reference configuration. The expression $\nabtil \theta$ denotes the temperature gradient with respect to the reference configuration.

To specify the general structure of the adhesive's material model, an ansatz for the Helmholtz free energy function $\hat{\psi} = \rhotil \psi$ per unit volume is introduced. It is additively decomposed into three parts according to
\begin{equation}
\label{eq:free_energy_allg}
  \hat{\psi} 
     = \hat{\psi}_{G}\Big(\Cg,\theta,z\Big) 
     + \hat{\psi}_V\Big(J_M\Big)
     + \hat{\psi}_{\theta C}\Big(\theta,q\Big)\ .
\end{equation}
Therein, $\hat{\psi}_{G}$ represents the stored energy as a result of isochoric deformations described by $\Cg$. Furthermore, this part depends on the current temperature~$\theta$ and on an additional material function $z$ which reflects some process dependencies.\footnote{ The isochoric part of the free energy may be extended by additional internal variables. This would be necessary if, for example, models of multiplicative viscoelasticity or viscoplasticity are employed (cf. \cite{Landgraf_Ihlemann_2011}).} The second contribution of Eq. \eqref{eq:free_energy_allg} describes the material response due to pure mechanical volume changes and only depends on the volume ratio $J_M$ of the mechanical deformation. The remaining part of Eq.~\eqref{eq:free_energy_allg} defines the thermochemically stored energy $\hat{\psi}_{\theta C} = \hat{\psi}_{\theta C}(\theta,q)$ of the material which is a function of the temperature $\theta$ and the degree of cure $q$. It is attributed to an amount of energy, which is initially stored in the material and which gets released due to the exothermic chemical process during curing. Furthermore, it describes the energy storage related to varying temperatures.

In Eq.~\eqref{eq:free_energy_allg}, the variables $q$ and $z$ are treated as internal variables. Thus, they are prescribed by evolution equations which are defined in general form by 
\begin{equation}
\label{eq:qdot}
  \dot{q} = f_q(q,\theta,t) \ge 0 \ , \quad q(t=0) = q_0 \ ,
\end{equation}
\begin{equation}
\label{eq:zdot}
  \dot{z} = f_z(q,\theta,t) \ge 0 \ , \quad z(t=0) = z_0 \ .
\end{equation}
Therein, $q_0$ and $z_0$ are appropriate initial conditions. It can be seen that both variables are monotonically increasing. More details and specific constitutive functions will be provided in Section \ref{sec:Curing_model_constitutive_functions}. 

To evaluate the ansatz \eqref{eq:free_energy_allg} within the Clausius-Duhem inequality \eqref{eq:CDU}, the rate of the Helmholtz free energy function $\hat{\psi}$ has to be calculated. Taking into account all dependencies of Eq.~\eqref{eq:free_energy_allg}, its rate reads as 
\begin{equation}
\label{eq:free_energy_derivative}
\begin{array}{lcl}
  \dot{\hat{\psi}} 
     &=&  \left[ \dfrac{\partial \hat{\psi}_{G}}{\partial \theta}
              +\dfrac{\partial \hat{\psi}_{\theta C}}{\partial \theta}
        \right] \, \dot{\theta}
    \, + \, \dfrac{\partial \hat{\psi}_{G}}{\partial \Cg} \ppkt \Cg!^\SLdreieck \ + 
     \\[4mm]
   & & \ \
    \, + \, \dfrac{\partial \hat{\psi}_{V}}{\partial J_M} \dot{J}_M
    \, + \, \dfrac{\partial \hat{\psi}_{\theta C}}{\partial q} \, \dot{q}
    \, + \, \dfrac{\partial \hat{\psi}_{G}}{\partial z} \dot{z} \ .
\end{array}
\end{equation}
A substitution of expressions \eqref{eq:CMdot}~-~\eqref{eq:Cgdot} and \eqref{eq:free_energy_derivative} into the Clausius-Duhem inequality \eqref{eq:CDU} yields the dissipation inequality
\begin{equation}
\label{eq:dissip_ineq}
\begin{array}{rcl}
  \left\{ 
     \dfrac{1}{2} \, \Ttil 
     - \left[ \dfrac{\partial \hat{\psi}_{G}}{\partial \Cg}\cdot \Cg\right]'\cdot\C^\inv + \dfrac{J_M}{2} \,\dfrac{\partial \hat{\psi}_V}{\partial J_M}\,\C^\inv 
  \right\} \ppkt \C!^\SLdreieck  \ + \ \ \\[5mm]
   - \left\{
       \rhotil \, \eta 
       + \dfrac{\partial \hat{\psi}_{\theta C}}{\partial \theta} 
       + \dfrac{\partial \hat{\psi}_{G}}{\partial \theta} 
       - \left( \dfrac{J_M}{J_{\theta C}}\,
                 \dfrac{\partial J_{\theta C}}{\partial \theta}\,
                \dfrac{\partial \hat{\psi}_V}{\partial J_M}
          \right) \right\}\, \dot{\theta} \ + \ \ \\[3mm]
   - \left\{
       \dfrac{\partial \hat{\psi}_{\theta C}}{\partial q} 
       - \left( \dfrac{J_M}{J_{\theta C}}\,
                 \dfrac{\partial J_{\theta C}}{\partial q}\,
                \dfrac{\partial \hat{\psi}_V}{\partial J_M}
          \right) \right\}\, \dot{q} \ + \ \ \\[3mm]
    - \dfrac{\partial \hat{\psi}_{G}}{\partial z} \dot{z}
    - \dfrac{1}{\theta} \, \qtil \cdot \nabtil \theta \ge 0 \ ,
\end{array}
\end{equation}
which has to be satisfied for arbitrary thermomechanical processes. Following the standard methods for the evaluation of \eqref{eq:dissip_ineq} (cf. \cite{Haupt_2002}), it is firstly stated that the terms in brackets in front of the $\C!^\SLdreieck$ and $\dot{\theta}$ have to be zero. This yields the potential relations for the $2^{\rm nd}$ Piola-Kirchhoff stress tensor
\begin{equation}
\label{eq:Ttil_allg}
 \Ttil 
     = 2\,\left[ \dfrac{\partial \hat{\psi}_{G}}{\partial \Cg}\cdot \Cg\right]'\cdot\C^\inv + J_M\,\dfrac{\partial \hat{\psi}_V}{\partial J_M}\,\C^\inv \ ,
\end{equation}
and the entropy 
\begin{equation}
\label{eq:eta_allg}
 \rhotil \, \eta 
       = - \dfrac{\partial \hat{\psi}_{\theta C}}{\partial \theta} 
          - \dfrac{\partial \hat{\psi}_{G}}{\partial \theta} 
          + \left( \dfrac{1}{J_{\theta C}}\,
                 \dfrac{\partial J_{\theta C}}{\partial \theta}\,
                \dfrac{\partial \hat{\psi}_V}{\partial J_M}
                J_M
          \right) \ .
\end{equation} 

Next it is assumed, that each of the remaining terms of inequality \eqref{eq:dissip_ineq} has to be non-negative which is a sufficient but not necessary condition. The non-negativity of the last term of inequality \eqref{eq:dissip_ineq} is complied by Fourier’s law. Formulated on the reference configuration it reads as
\begin{equation}
\label{eq:Fourier}
  \qtil = - \kappa \, J \, \C^\inv \cdot \nabtil \theta\ .
\end{equation} 
Here, $\kappa \ge 0$ is the thermal conductivity. Furthermore, taking into account the properties $\dot{q} \ge 0$ and $\dot{z} \ge 0$ (see Eqs. \eqref{eq:qdot} and \eqref{eq:zdot}), the final two restrictions read as
\begin{equation}
\label{eq:dissi_qdot}
 - \left\{
       \dfrac{\partial \hat{\psi}_{\theta C}}{\partial q} 
       - \left( \dfrac{1}{J_{\theta C}}\,
                 \dfrac{\partial J_{\theta C}}{\partial q}\,
                \dfrac{\partial \hat{\psi}_V}{\partial J_M}
                J_M
          \right) \right\} \ \ge 0 \ ,
\end{equation} 
\begin{equation}
\label{eq:dissi_zdot}
    - \dfrac{\partial \hat{\psi}_{G}}{\partial z} \ge 0 \ .
\end{equation} 
These conditions cannot be evaluated in general form. However, for the case of the concretized material model presented in Section~\ref{sec:Curing_model_constitutive_functions}, the thermodynamic consistency is proved (cf. Section~\ref{sec:Constitutive_functions_Process_consistency}).

%
\section{Application to an epoxy based adhesive}
\label{sec:Curing_model_constitutive_functions}

In this section, the general modelling framework is specified to simulate the material behaviour of one specific class of  adhesives. More precisely, the two-part epoxy based structural adhesive \textit{DP410}$^{\rm TM}$ provided by \textit{3M Scotch-Weld}$^{\rm TM}$ is modelled \cite{DP410_2003}. The adhesive is composed by mixing of two paste-like components. Afterwards, the mixture cures to a solid without any further initiation. In particular, curing takes place at room temperature such that no heating is necessary. The fully cured material can be applied within a temperature range of $-55^\circ C$ to $80^\circ C$ and the glass transition temperature is about $50^\circ C$. Furthermore, the mass density is approximately $1.1 \, \rm g/cm^3$ \cite{DP410_2003}. In the following, different aspects of the model specifications are addressed.

\subsection{Degree of cure}
\label{sec:Constitutive_functions_Degree_of_Cure}

First of all, the curing process is examined in more detail. In analogy to the procedures described in \cite{Kolmeder_Lion_2010} and \cite{Lion_Yagimli_2008}, the curing process has been measured by Differential Scanning Calorimetry (DSC) experiments. Thus, it is assumed that the curing process can completely be determined by the exothermic reaction during the chemical process. According to Halley and Mackay \cite{Halley_Mackay_1996}, different phe\-no\-me\-no\-lo\-gi\-cal models can be applied to simulate the curing processes of epoxy based materials. In this work, the so called $n$-th-order model (model of reaction order $n$)  is employed. In view of Eq.~\eqref{eq:qdot}, this specific ansatz is expressed by
\begin{equation}
\label{eq:q_ansatz}
   \dot{q} =  f_q(q,\theta,t) = K_1(\theta) \cdot\Big(1-q\Big)^n \cdot f_D(q,\theta) \ .
\end{equation}
Therein, $n$ is a constant material parameter and $K_1(\theta)$ is a temperature dependent thermal activation function which is constituted by the Arrhenius ansatz  (cf. \cite{Halley_Mackay_1996})
\begin{equation}
\label{eq:q_Kfunc}
  K_1(\theta) = K_{10} \, {\rm exp}\left[-\frac{E_1}{R\,\theta}\right] \ ,
\end{equation}
where $K_{10}$ and $E_1$ are constant material parameters and \mbox{$R = 8.3144 \, \rm J/(mol\,K)$} is the universal gas constant. To account for diffusion controlled curing, which takes place at temperatures below the glass transition temperature, the ansatz \eqref{eq:q_ansatz} includes an empirical diffusion factor  $f_D(q,\theta)$ which, according to Fournier \etal \cite{Fournier_Etal_1996}, reads as
\begin{equation}
\label{eq:q_diffusion}
  f_D(q,\theta) = \dfrac{2}{1+{\rm exp}\left[\frac{q-q_{end}(\theta)}{b}\right]}-1 \ .
\end{equation}
Here, $b$ is another constant material parameter and $q_{end}$ is the maximum degree of cure, which can be attained at a certain temperature $\theta$. To evaluate the maximum attainable degree of cure, typically the DiBenedetto equation is adopted \cite{DiBenedetto_1987,Kolmeder_Lion_2010,Pascault_Williams_1990}:
\begin{equation}
\label{eq:q_diBenedetto}
  \dfrac{T_g(q)-T_{g,0}}{T_{g,1}-T_{g,0}} = \dfrac{\lambda \, q}{1- (1-\lambda) \, q} \ .
\end{equation}
Therein, $T_g(q)$ is the glass transition temperature as a function of the degree of cure and $T_{g,0}$ and $T_{g,1}$ are the glass transition temperatures at degree of cure $q=0$ and $q=1$, respectively. Furthermore, $\lambda$ is a constant material parameter. In order to calculate the maximum attainable degree of cure at certain isothermal curing temperatures, Eq.~\eqref{eq:q_diBenedetto} has to be solved for $q$ as follows
\begin{equation}
\label{eq:q_qend}
  q_{end}(\theta) = \dfrac{f_T(\theta)}{f_T(\theta)-\lambda \, f_T(\theta) + \lambda} \ .
\end{equation}
Here, an abbreviation $f_T(\theta)$ has been introduced 
\begin{equation}
\label{eq:q_fTheta}
  f_T(\theta) = \dfrac{\theta + \Delta T - T_{g,0}}{T_{g,1}-T_{g,0}} \ .
\end{equation}
In Eq.~\eqref{eq:q_fTheta} the assumption $T_g(q) = \theta + \Delta T$ has been employed. Therein, $\Delta T$ denotes the difference between the glass transition temperature $T_g(q)$ attainable at specific isothermal curing temperatures, and the curing temperature~$\theta$ itself.

\begin{figure}[ht]
	\centering
  \includegraphics[width=0.45\textwidth]{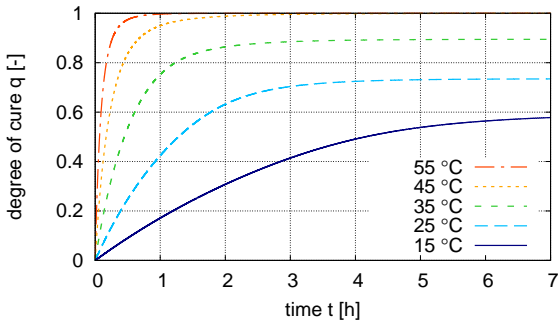}
	\caption{Evolution of the degree of cure $q$ for different temperatures $\theta$}
	\label{fig:curing}
\end{figure}

The material parameters of the model \eqref{eq:q_ansatz} - \eqref{eq:q_fTheta} have been identified using the DSC measurements. The corresponding values are listed in Table~\ref{tab:MatPar_Cure}. Moreover, the phenomenological behaviour of this model is depicted in Fig. \ref{fig:curing} for different temperatures.

\begin{table}[ht]
 \centering
  \caption{Material parameters for Eqs.~\eqref{eq:q_ansatz} - \eqref{eq:q_fTheta}} 
  {\begin{tabular}{p{1.6cm}p{2.0cm}p{1.6cm}p{1.4cm}}
    \hline & & & \\[-3mm]
    parameter & value & parameter & value\\ 
    \hline & & & \\[-3mm]
        $\ \ $ $K_{10}$    & $1.608\cdot10^{10} \rm$  
     & $\ \ $ $T_{g,1}$   & $324.85 \ \rm K$   \\
        $\ \ $ $E_1$        & $79835 \ \rm J/mol$         
     & $\ \ $ $T_{g,0}$   & $234.35 \ \rm K$   \\  
        $\ \ $ $b$            & $0.057$                         
     & $\ \ $ $\Delta T$  & $11 \ \rm K$   \\
        $\ \ $ $n$            & $1.217$                          
     & $\ \ $ $\lambda$  & $1.7$   \\ 
    \hline
  \end{tabular}}
  \label{tab:MatPar_Cure}
\end{table}  

\subsection{Heat expansion and chemical shrinkage}
\label{sec:Constitutive_functions_Volume}

Next, the thermochemical volume change due to chemical shrinkage and heat expansion processes is specified. To this end, an idealized model with the following ansatz has been chosen:
\begin{equation}
\label{eq:phi_thetaC}
  \varphi_{\theta C}(\theta, q) = {\rm exp}\left[ \alpha_\theta \, \big(\,\theta - \thetatil\,\big) + \beta_q \, q\right] \ .
\end{equation}
Therein, $\alpha_\theta$ is a volumetric heat expansion coefficient and $\beta_q$ is the maximum volumetric chemical shrinkage. According to first measurement results, the material parameters have been set to the values listed in Table~\ref{tab:MatPar_Volume}.

\begin{table}[ht]
 \centering
  \caption{Material parameters for Eq.~\eqref{eq:phi_thetaC}} 
  {\begin{tabular}{p{1.6cm}p{2.0cm}p{1.6cm}p{1.4cm}}
    \hline & & & \\[-3mm]
    parameter & value & parameter & value\\
    \hline & & & \\[-3mm]
        $\ \ $ $\thetatil$           & $295 \ \rm K$ 
    &                                   &   \\
        $\ \ $ $\alpha_\theta$  & $5\cdot10^{-4} \ \rm K^\inv$ 
    &  $\ \ $ $\beta_q$          & $-0.05$    \\[1mm]
    \hline
  \end{tabular}}
  \label{tab:MatPar_Volume}
\end{table}  

\subsection{Free energy and stresses}
\label{sec:Constitutive_functions_Free_Energy}

\begin{figure}[ht]
	\centering
  \includegraphics[width=0.25\textwidth]{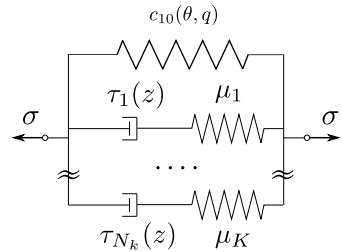}
	\caption{Rheological model of a nonlinear spring connected in parallel to multiple Maxwell elements}
	\label{fig:rheolog_model}
\end{figure}

To complete the curing model, the mechanical parts of the free energy function \eqref{eq:free_energy_allg} and thus the corresponding stress strain relationships have to be specified. Firstly, the mechanical response due to isochoric deformations is considered. It is described by the free energy contribution $\hat{\psi}_{G}(\Cg,\theta,z)$ in Eq.~\eqref{eq:free_energy_allg}. This part is modelled by a combination of a finite strain pseudo-elasticity with temperature and degree of cure dependent stiffness and a sum of multiple Maxwell elements, each including process dependencies described by the material function $z(t)$ (cf. Eq.~\eqref{eq:zdot}). Fig.~\ref{fig:rheolog_model} illustrates this model by means of a one-dimensional rheological representation. 

The corresponding free energy  $\hat{\psi}_{G}$  is constituted as a sum of contributions related to a pseudo-elastic part $\hat{\psi}_{el}$ and $N_k$ Maxwell elements, each denoted by $\hat{\psi}_{ve,k}$:
\begin{equation}
\label{eq:psi_G}
  \hat{\psi}_{G}(\Cg,\theta,z)
     = \hat{\psi}_{el}\Big(\Cg,\theta\Big) 
     + \sum_{k=1}^{N_k} \hat{\psi}_{ve,k}\Big(\Cg,z\Big) \ .
\end{equation}
The pseudo-elastic part is modelled by an ansatz proposed by Lion and Johlitz \cite{Lion_Johlitz_2012}. It takes the form
\begin{equation}
\label{eq:psi_ela}
\begin{array}{ll}
  \hat{\psi}_{el}\Big(\Cg,\theta\Big) = \Ten2 Q \ppkt\Cg , 
\end{array}
\end{equation}
where the tensor $\Ten2 Q$ is given by
\begin{equation}
\label{eq:psi_ela_II}
\begin{array}{ll}
  \displaystyle
  \Ten2 Q = - \int\limits_{-\infty}^{t}2\,c_{10}\Big(\theta(t),q(s)\Big)\,\Bigg(\dfrac{{\rm d}}{{\rm d}s}\C!^\SLstrich^\inv(s)\Bigg)\ {\rm d} s\ .
\end{array}
\end{equation}
Therein, a stiffness function $c_{10}(\theta,q)$ is introduced. It takes into account the dependency on the curing history described by the degree of cure $q(s)$ ($s$ is the integration variable). Furthermore, a dependency on the current temperature $\theta(t)$ is included. The stiffness function exhibits the properties
\begin{equation}
\label{eq:psi_ela_stiffness}
  c_{10}\Big(\theta,q\Big)
    \ge 0 \ , \quad
  \dfrac{\partial  }{\partial q} \, c_{10}\Big(\theta,q\Big) \ge 0   \ .
\end{equation}
A specific ansatz will be provided in Section \ref{sec:Constitutive_functions_Process_dependency}. The free energy $\hat{\psi}_{ve,k}$ of one single Maxwell element (see Eq.~\eqref{eq:psi_G}) is modelled according to an ansatz proposed by Haupt and Lion \cite{Haupt_Lion_2002}:
\begin{equation}
\label{eq:psi_visc_single}
  \displaystyle
  \hat{\psi}_{ve,k} 
     = \left\{ - \int\limits_{-\infty}^{z}G_k(z-s)\,\Bigg(\dfrac{{\rm d}}{{\rm d}s}\C!^\SLstrich^\inv(s)\Bigg)\ {\rm d} s\right\}\ppkt\Cg \ .
\end{equation}
Therein, $G_k(z-s)$ is a relaxation function which is constituted by
\begin{equation}
\label{eq:kernel}
  \displaystyle
  G_k(z-s)
     = 2 \, \mu_k \, {\rm e}^{-\frac{z-s}{\tau_k}} \ .
\end{equation}
Therein, ${\rm e}$ denotes the Euler's number and the parameters $\mu_k$ and $\tau_k$ are the stiffness and the relaxation time, respectively, for the $k$-th Maxwell element (cf. Fig.~\ref{fig:rheolog_model}). Note that Eq.~\eqref{eq:psi_visc_single} is formulated with respect to the material function $z(t)$ instead of the physical time $t$. The variable $z(t)$ is also referred to as intrinsic time scale and is governed by an evolution equation which has been introduced in general form by Eq.~\eqref{eq:zdot}. Different sources of process dependencies may be defined by choosing appropriate constitutive functions for this evolution equation. However, only temperature and degree of cure dependent behaviour is assumed in this work. A specific ansatz for Eq.~\eqref{eq:zdot} is provided in Section~\ref{sec:Constitutive_functions_Process_dependency}.

To complete the mechanical part of the free energy function \eqref{eq:free_energy_allg}, the volumetric stress response described by $\hat{\psi}_V$ has to be constituted. This contribution is assumed to be pure elastic and is described by the ansatz
\begin{equation}
\label{eq:psi_vol}
  \displaystyle
  \hat{\psi}_V  = \dfrac{K}{2} \, \Big(J_M - 1\Big)^2 \ .
\end{equation}
Therein, $K>0$ is the bulk modulus. 

Finally, the $2^{\rm nd}$ Piola-Kirchhoff stress tensor $\Ttil$ is calculated by evaluation of Eq.~\eqref{eq:Ttil_allg} in combination with the specific constitutive relations \eqref{eq:psi_G} - \eqref{eq:psi_vol}. The resulting contributions to the $2^{\rm nd}$ Piola-Kirchhoff stress tensor are summarized in Eqs.~\eqref{eq:stress_sum} - \eqref{eq:stress_visc_k}. More information on the process dependent material functions $c_{10}(\theta,q)$ and~$\dot{z}$ and a summary of specific values for material parameters are provided in Section \ref{sec:Constitutive_functions_Process_dependency}. 

\begin{center}
\hrule \footnotesize
\nopagebreak
\vspace{1ex}
\begin{eqnarray}
\omit\rlap{\text{Total $2^{\rm nd}$ PK stress}} \nonumber \\[1mm]
\quad  \Ttil \ \ \,\,\,  \;
       &=& \ \Ttil_{V} +\displaystyle  \Big(\Ttil_{G}\cdot\Cg\Big)'\cdot\C^\inv  \label{eq:stress_sum} \\[1mm]
\omit\rlap{\text{Volumetric part}} \nonumber \\[1mm]
\quad \Ttil_{V} \,\,\,\,
       &=& \ K \, J_M \, (J_M - 1 ) \, \C^\inv \label{eq:stress_vol}  \\[1mm]
\omit\rlap{\text{Isochoric part}} \nonumber \\[1mm]
\quad \Ttil_{G} \,\,\,\,
       &=&  \ \Ttil_{el} + \Ttil_{ve} \label{eq:stress_iso} \\[1mm]
\omit\rlap{\text{Pseudo-elastic part}} \nonumber \\[1mm]
\quad \Ttil_{el}\,\,\,\,
       &=&  - \int\limits_{-\infty}^{t}2\,c_{10}\Big(\theta(t),q(s)\Big)\,
                  \Bigg(\frac{{\rm d}}{{\rm d}s}\C!^\SLstrich^\inv(s)\Bigg)\ {\rm d} s  \label{eq:stress_ela} \\[1mm]
\omit\rlap{\text{Viscoelastic part}} \nonumber \\[1mm]
\quad \Ttil_{ve}\,\,\,
       &=&  \sum_{k=1}^{N_k} \ \Ttil_{ve,k} \label{eq:stress_visc} \\[1mm]
\quad \Ttil_{ve,k}
       &=&  - \int\limits_{-\infty}^{z(t)}2 \mu_k\,{\rm e}^{-\frac{z(t)-s}{\tau_k}} 
                  \bigg(\dfrac{\rm d}{{\rm d}s}\Cg^\inv(s)\bigg) \,{\rm d}s \label{eq:stress_visc_k} 
\end{eqnarray}
\nopagebreak
\hrule 
\end{center}

\subsection{Process dependencies of mechanical properties}
\label{sec:Constitutive_functions_Process_dependency}

In Eq.~\eqref{eq:psi_ela} a stiffness parameter $c_{10}(\theta,q)$ has been introduced which includes dependencies on the temperature $\theta$ and the degree of cure $q$. The specific ansatz used in this paper consists of a separation of the different physical processes
\begin{equation}
\label{eq:c10_ansatz}
   c_{10}(\theta,q) = c_{10,0}\,f_{c\theta}(\theta)\,f_{cq}(q) \ .
\end{equation}
Therein, $c_{10,0}$ is a constant stiffness parameter which equals the half shear modulus $G$ in small strain shear experiments. Furthermore, $f_{c\theta}(\theta)$ and $f_{cq}(q)$ are normalized functions representing the temperature and degree of cure dependencies, respectively. The normalization is accomplished in a way, such that the corresponding values of both functions range from $0$ to $1$. 

\begin{figure}[ht]
   \centering
   \includegraphics[width=0.45\textwidth]{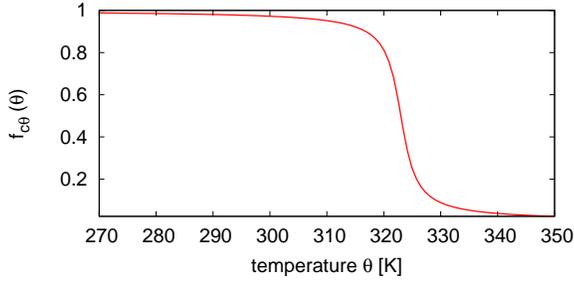}
   \caption{Temperature dependency $f_{c\theta}(\theta)$ of the equilibrium stiffness}
   \label{fig:ela_temp_func}
\end{figure}

\begin{figure}[ht]
   \centering
	 \includegraphics[width=0.45\textwidth]{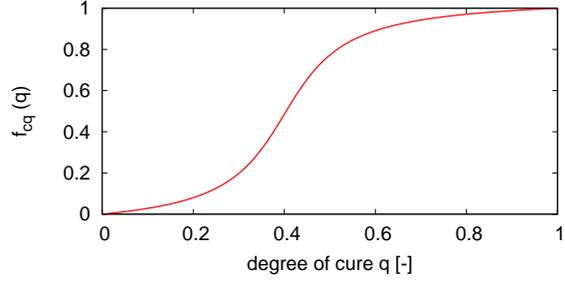}
   \caption{Degree of cure dependency $f_{cq}(q)$ of the 
               equilibrium stiffness}
	\label{fig:ela_cure_func}
\end{figure}

For the representation of the temperature dependency of the fully cured material, the normalized function $f_{c\theta}(\theta)$ is constituted by the ansatz
\begin{equation}
\label{eq:c10_fTheta}
  f_{c\theta}\big(\theta\big) = \dfrac{1}{\pi}\Bigg\{{\rm atan}\Big[ a_{c\theta} \cdot\big(\theta - T_{g,1}\big)\Big]  + \dfrac{\pi}{2}\Bigg\} \ .
\end{equation}
It takes into account the major part of stiffness change near the glass transition temperature $T_{g,1}$ of the cured material  (cf. Table~\ref{tab:MatPar_Cure}). An additional material parameter $a_{c\theta}$ enables one to adjust the specific shape of the function. Fig.~\ref{fig:ela_temp_func} illustrates the phenomenology of Eq.~\eqref{eq:c10_fTheta}. The chosen material parameter $a_{c\theta}$ is listed in Table~\ref{tab:MatPar_Ela}.
The second normalized function $f_{cq}(q)$ of Eq.~\eqref{eq:c10_ansatz} represents the change in stiffness due to the curing process and thus only depends on the degree of cure. The chosen ansatz reads as
\begin{equation}
\label{eq:c10_gC}
  f_{cq}\big(q\big) = \dfrac{1}{d_{cq}}\Bigg\{{\rm atan}\Big[ a_{cq} \cdot\big(q - b_{cq}\big)\Big]  + c_{cq}\Bigg\} \ .
\end{equation}
Therein, $a_{cq}$ and $b_{cq}$ are material parameters and the variables $c_{cq}$ and $d_{cq}$ are evaluated in a way such that the conditions \mbox{$f_{cq}(q=0) = 0$} and \mbox{$f_{cq}(q=1)=1$} hold. An evaluation of both conditions yields the expressions
\begin{equation}
\label{eq:c10_gC_II}
  c_{cq}= - {\rm atan}\Big[ - a_{cq}\cdot b_{cq}   \Big], \quad
\end{equation}
\begin{equation}
\label{eq:c10_gC_III}
  d_{cq}=   {\rm atan}\Big[ a_{cq}\cdot (1-b_{cq})\Big] + c_{cq} \ .
\end{equation}
The course of the function $f_{cq}(q)$ is depicted in Fig.~\ref{fig:ela_cure_func}. The material parameters $a_{cq}$ and $b_{cq}$ used for illustration are listed in Table~\ref{tab:MatPar_Ela}.

\begin{table}[ht]
 \centering
  \caption{Material parameters for Eqs.~\eqref{eq:stress_vol}, \eqref{eq:stress_ela} 
                and \eqref{eq:c10_ansatz} - \eqref{eq:c10_gC}} 
  {\begin{tabular}{p{1.6cm}p{2.0cm}p{1.6cm}p{1.4cm}}
    \hline& & & \\[-3mm]
    parameter & value & parameter & value\\ 
    \hline& & & \\[-3mm]
          $\ \ $ $K$               & $5000 \, \rm MPa$     
      &  $\ \ $ $c_{10,0}$     & $500 \, \rm MPa$     \\[1mm]
          $\ \ $ $a_{c\theta}$  & $- 0.5 \, \rm K^\inv$    
      &  $\ \ $ $a_{cq}$        & $10$   \\[1mm]
                                        &
      & $\ \ $ $b_{cq}$         & $0.4 $     \\  [1mm]
    \hline
  \end{tabular}}
  \label{tab:MatPar_Ela}
\end{table}  

Finally, the process dependency of the viscoelastic part of Eq.~\eqref{eq:psi_G} is considered. A set of several Maxwell elements has been modelled to capture the materials viscoelastic behaviour. As presented in Section \ref{sec:Constitutive_functions_Free_Energy}, the process dependency is accommodated by the intrinsic time scale $z(t)$. The following ansatz for the evolution equation \eqref{eq:zdot} has been chosen
\begin{equation}
\label{eq:zdot_ansatz}
  \displaystyle
  \dot{z}  
      \,=\, f_z(q,\theta,t) = 10^{f_{z\theta}(\theta)}\cdot10^{f_{zq}(q)} \ .
\end{equation}
In accordance to the pseudo-elastic stiffness \eqref{eq:c10_ansatz}, the dependencies on the temperature and the degree of cure have been separated in Eq.~\eqref{eq:zdot_ansatz}. The constitutive equations for both functions $f_{z\theta}(\theta)$ and $ f_{zq}(q) $ are
\begin{align}
 f_{z\theta}(\theta) 
      &= \dfrac{a_z}{\pi} \, {\rm atan}\Big[b_z\cdot[\theta-T_{g,1}]\Big]
           +\dfrac{\pi}{2} \ ,\label{eq:zdot_ansatz_function_theta}
  \\[2mm]
  f_{zq}(q) 
      &= c_z\cdot(1-q^{n_z}) \ .
\label{eq:zdot_ansatz_function_q}
\end{align}
The material parameters belonging to the viscoelastic part of the model are listed in Table~\ref{tab:MatPar_Visc}. 

\begin{table}[ht]
 \centering
  \caption{Material parameters for Eqs.~\eqref{eq:stress_visc}, \eqref{eq:stress_visc_k}, \eqref{eq:zdot_ansatz}  - \eqref{eq:zdot_ansatz_function_q}} 
  {\begin{tabular}{p{1.6cm}p{2.0cm}p{1.6cm}p{1.4cm}}
    \hline & & & \\[-3mm]
    parameter & value & parameter & value\\ 
    \hline & & & \\[-3mm]
        $\ \ $ $N_k$            & $7$                       
    &  $\ \ $ $\mu_{2-7}$  & $  5 \, \rm MPa$    \\
        $\ \ $ $\mu_{1}$     & $75 \, \rm MPa$    
    &  $\ \ $ $\tau_{2-4}$  & $10^{k-4} \, \rm s$  \\  
        $\ \ $ $\tau_1$        & $10 \, \rm s$   
    &  $\ \ $ $\tau_{5-7}$  & $10^{k-3} \, \rm s$  \\[1mm]
        $\ \ $ $a_z$            & $6.0$                     
    &  $\ \ $ $b_z$            & $0.05 \ \rm K^\inv$    \\ 
        $\ \ $ $c_z$            & $5.0$                     
    &  $\ \ $ $n_z$            & $0.6$    \\ [1mm]
    \hline
  \end{tabular}}
  \label{tab:MatPar_Visc}
\end{table}  

%
\section{Aspects of finite element implementation}
\label{sec:FEM_implementation}

The constitutive model for the representation of curing phenomena in adhesives has been implemented into the finite element software \textit{ANSYS}$^{\rm TM}$. In this section, the numerical integration of constitutive equations as well as the derivation of \textit{ANSYS}$^{\rm TM}$ specific stress and material tangent measures are summarized. Additionally, a new algorithm is introduced which suppresses undesired initial volume changes. Those volume changes may result from thermal expansion and chemical shrinkage when initial values for the temperature and the degree of cure differ from their reference values.

\subsection{Numerical integration}
\label{sec:FEM_implementation_integration}

For the numerical integration of constitutive equations, a typical time interval ($t_n$, $t_{n+1}$) with $\Delta t = t_{n+1} - t_n > 0$ is considered. Within this time step, the $2^{\rm nd}$ Piola-Kirchhoff stress \eqref{eq:stress_sum} has to be computed. Thus,  
\begin{equation}
\label{eq:stresses_incremental}
  \indLT{n+1}\Ttil
     = \indLT{n+1}\Ttil_{V}
       + \Big(\indLT{n+1}\Ttil_{G}\cdot\indLT{n+1}\Cg\,\Big)'\cdot\indLT{n+1}\C^\inv 
\end{equation}
has to be solved.  Here, values at time instance  $t_{n+1}$ are denoted by $\indLT{n+1}(\cdot)$. Accordingly, values at time instance $t_n$ are indicated by $\indLT{n}(\cdot)$ and values that are defined at the midpoint $t_n + \frac{\Delta t }{2}$ are represented by $\indLT{n/2}(\cdot)$. Within a single time step $\Delta t$, it is assumed that the deformation gradients $\indLT{n}\F$ and $\indLT{n+1}\F$ as well as the temperatures $\indLT{n}\theta$ and $\indLT{n+1}\theta$ are known. Furthermore, internal variables of the preceding time step are given. 

Firstly, the solution for the degree of cure  $\indLT{n+1}q$ is obtained by numerical integration of the evolution Eq.~\eqref{eq:q_ansatz}. Here, Euler backward method (Euler implicit) is employed. A formulation of Eq.~\eqref{eq:q_ansatz} for time instance $t_{n+1}$ and a substitution of the approximation $\indLT{n+1}{\dot{q}} \approx \frac{1}{\Delta t}(\indLT{n+1}q-\indLT{n}q)$ yields 
\begin{equation}
\label{eq:q_EBM}
  \indLT{n+1}q = \indLT{n}q + \Delta t \ f_q(\indLT{n+1}q,\indLT{n+1}\theta) \ ,
\end{equation}
which is a nonlinear equation with respect to the solution $\indLT{n+1}q$. It is computed by application of Newton's method. Additionally, the degree of cure $\indLT{n/2}q$ at time instance $t = t_n + \frac{\Delta t}{2}$ has to be computed as well. In consideration of the temperature $\indLT{n/2}\theta = \frac{1}{2}(\indLT{n+1}\theta + \indLT{n}\theta)$, the value $\indLT{n/2}q$ is obtained according to \eqref{eq:q_EBM} by
\begin{equation}
\label{eq:q_EBM_n2}
  \indLT{n/2}q = \indLT{n}q + \dfrac{\Delta t}{2} \ f_q(\indLT{n/2}q,\indLT{n/2}\theta) \ .
\end{equation}
Next, the computation of the intrinsic time scale $\indLT{n+1}z$ is considered. Since this variable is governed by an evolution equation \eqref{eq:zdot_ansatz} which cannot be solved in closed form, a numerical scheme has to be applied as well. By analogy with Eq.~\eqref{eq:q_EBM}, Euler backward method and the approximation $\indLT{n+1}{\dot{z}} \approx \frac{1}{\Delta t}(\indLT{n+1}z-\indLT{n}z)$ are adopted
\begin{equation}
\label{eq:z_EBM}
  \indLT{n+1}z = \indLT{n}z + \Delta t \ f_z(\indLT{n+1}q,\indLT{n+1}\theta) \ .
\end{equation}
In contrast to Eqs.~\eqref{eq:q_EBM} and \eqref{eq:q_EBM_n2}, this relation can directly be evaluated. Thus, no iterative procedure has to be applied.

Next, the calculation of the stresses Eq.~\eqref{eq:stresses_incremental} is considered. While $\Ttil_V$ is already completely determined (see Eq.~\eqref{eq:stress_vol}), the stress contribution $\Ttil_G$ has to be regarded in more detail. According to Eq. \eqref{eq:stress_iso}, it includes a pseudo-elastic part $\indLT{n+1}\Ttil_{el}$ and a sum of several viscoelastic parts summarized by $\indLT{n+1}\Ttil_{ve}$. 

To compute the pseudo-elastic stress tensor, firstly Eq.~\eqref{eq:stress_ela} has to be formulated for time instance $t_{n+1}$:
\begin{equation}
\label{eq:Tel_incemental}
\displaystyle  \indLT{n+1}\Ttil_{el}
          = - \int\limits_{-\infty}^{t_{n+1}}2\,c_{10}( \indLT{n+1}\theta, q(s) ) \, \Bigg(\dfrac{{\rm d}}{{\rm d}s}\C!^\SLstrich^\inv(s)\Bigg)\ {\rm d} s \ .
\end{equation}
Note that the temperature $\indLT{n+1}\theta$ does not depend on the integration variable $s$ but only on time instance $t_{n+1}$. 

After substituting the ansatz \eqref{eq:c10_ansatz} for the stiffness $c_{10}(\theta,q)$, all values that do not depend on the integration variable $s$ are excluded from the integral such that the pseudo-elastic stress can be rewritten by 
\begin{align}
   \displaystyle
   \indLT{n+1}\Ttil_{el}
          &= -  2\, c_{10,0} \, f_{c\theta}\big(\indLT{n+1}\theta\big) \  \indLT{n+1}\P_{el} \label{eq:Tel_incemental_abbrv} \ .
\end{align}
Here, $\indLT{n+1}\P_{el}$ is an abbreviation for the remaining integral 
\begin{align}
  \displaystyle
   \indLT{n+1}\P_{el}  
      &=     \int\limits_{-\infty}^{t_{n+1}} \, f_{cq}\big(q(s)\big) \, \Bigg(\dfrac{{\rm d}}{{\rm d}s}\C!^\SLstrich^\inv(s)\Bigg)\ {\rm d} s \ .
       \label{eq:Tel_incemental_reduced}
\end{align}
Next, this integral is split into two sub-integrals by substituting $t_{n+1} = t_n + \Delta t$ 
\begin{equation}
\label{eq:Tel_incemental_split}
\begin{array}{l}
   \displaystyle
   \indLT{n+1}\P_{el} 
             = \int\limits_{-\infty}^{t_{n}} \, f_{cq}\big(q(s)\big) \, \Bigg(\dfrac{{\rm d}}{{\rm d}s}\C!^\SLstrich^\inv(s)\Bigg)\ {\rm d} s \ +\\
    \qquad \qquad\qquad  \displaystyle
             +  \int\limits_{t_n}^{t_{n + \Delta t}} \, f_{cq}\big(q(s)\big) \, \Bigg(\dfrac{{\rm d}}{{\rm d}s}\C!^\SLstrich^\inv(s)\Bigg)\ {\rm d} s \ .
\end{array}
\end{equation}
The first term on the right-hand side of Eq.~\eqref{eq:Tel_incemental_split} is the solution of Eq. \eqref{eq:Tel_incemental_reduced} for time instance $t_n$. Thus, it equals $\indLT{n}\P_{el} $. The second term of Eq.~\eqref{eq:Tel_incemental_split} is computed numerically by the midpoint method 
\begin{equation}
\label{eq:Tel_incemental_midpoint}
\begin{array}{l}
 { \displaystyle
  \int\limits_{t_n}^{t_{n}+ \Delta t}} f_{cq}\big(q(s)\big) \, \Big(\frac{{\rm d}}{{\rm d}s}\C!^\SLstrich^\inv(s)\Big)\ {\rm d} s    \\
   \qquad \qquad \qquad
    \approx \Delta t \  f_{cq}\Big(\indLT{n/2}q\Big) \ \  \C!^\SLstrich!^\SLdreieck^\inv\Big|_{t_n + \frac{\Delta t}{2}} \ ,
\end{array}
\end{equation}
and the material time derivative of the isochoric inverse right Cauchy-Green tensor at time instance $t_n + \frac{\Delta t}{2}$ is approximated by
\begin{equation}
\label{eq:CG_approx}
\begin{array}{l}
  \Cg!^\triangle^\inv\Big|_{t_n + \frac{\Delta t}{2}}
  \approx
    \dfrac{1}{\Delta t} \Big(\indLT{n+1}\Cg^\inv  - \indLT{n}\Cg^\inv \Big) \ .
\end{array}   
\end{equation}
A substitution of Eqs.~\eqref{eq:Tel_incemental_midpoint} and \eqref{eq:CG_approx} into \eqref{eq:Tel_incemental_split} yields the incremental representation
\begin{equation}
\label{eq:Tel_incemental_solu}
\begin{array}{l}
   \displaystyle
   \indLT{n+1}\P_{el} 
             = \indLT{n}\P_{el} 
              +    f_{cq}\Big(\indLT{n/2}q\Big) \  \Big(\indLT{n+1}\Cg^\inv  - \indLT{n}\Cg^\inv \Big)\ ,
\end{array}
\end{equation}
and the pseudo-elastic stress $\indLT{n+1}\Ttil_{el}$ can be computed by Eq.~\eqref{eq:Tel_incemental_abbrv}. Finally, it remains to calculate the viscoelastic stress $\indLT{n+1}\Ttil_{ve}$. Note that the constitutive equations for the pseudo-elastic stress \eqref{eq:stress_ela} and the viscoelastic stress of one Maxwell element \eqref{eq:stress_visc_k} have a similar structure. Thus, the procedure for numerical integration is similar as well. However, the procedure described above has to be slightly adapted. Firstly, the stress contribution for one single Maxwell element \eqref{eq:stress_visc_k} is formulated for time instance $t_{n+1}$:
\begin{equation}
\label{eq:Tve_incemental}
  \indLT{n+1}\Ttil_{ve,k} 
      =   - \int\limits_{-\infty}^{\indLT{n+1}z}
                  2 \mu_k\,{\rm e}^{-\frac{\indLT{n+1}z-s}{\tau_k}} 
                  \bigg(\dfrac{\rm d}{{\rm d}s}\Cg^\inv(s)\bigg) \, {\rm d}s \ .
\end{equation}
Next, this integral is split into two sub-integrals which yields
\begin{equation}
\label{eq:Tve_incremental_split}
\begin{array}{l}
\displaystyle
  \indLT{n+1}\Ttil_{ve,k} 
      =   - \int\limits_{-\infty}^{\indLT{n}z}
                2 \mu_k\,{\rm e}^{-\frac{\indLT{n}z + \Delta z-s}{\tau_k}} 
                \bigg(\dfrac{\rm d}{{\rm d}s}\Cg^\inv(s)\bigg) \, {\rm d}s 
                \\[5mm]\displaystyle
    \qquad \qquad
          - \int\limits_{\indLT{n}z}^{\indLT{n}z + \Delta z}
                2 \mu_k\,{\rm e}^{-\frac{\indLT{n}z + \Delta z-s}{\tau_k}} 
                \bigg(\dfrac{\rm d}{{\rm d}s}\Cg^\inv(s)\bigg) \, {\rm d}s  .
\end{array}
\end{equation}
Note that in contrast to the calculation step \eqref{eq:Tel_incemental_split} here the intrinsic time scale $\indLT{n+1}z = \indLT{n}z + \Delta z$ has been substituted. The first integral on the right-hand side of Eq.~\eqref{eq:Tve_incremental_split} can be expressed by the solution $\indLT{n}\Ttil_{ve,k}$ as follows
\begin{equation}
\label{eq:Tve_incremental_first_term}
\begin{array}{l}
\displaystyle
-{\rm e}^{-\frac{\Delta z}{\tau_k}}
               \int\limits_{-\infty}^{\indLT{n}z}
                2 \mu_k\,{\rm e}^{-\frac{\indLT{n}z -s}{\tau_k}} 
                \bigg(\dfrac{\rm d}{{\rm d}s}\Cg^\inv(s)\bigg) \ {\rm d}s
                \\[5mm]\displaystyle
    \qquad \qquad
   = {\rm e}^{-\frac{\Delta z}{\tau_k}} \  \indLT{n}\Ttil_{ve,k}  \ .
  \end{array}
\end{equation}
Since the second term of Eq.~\eqref{eq:Tve_incremental_split} cannot be solved in closed form, a numerical procedure has to be applied. Here again, the midpoint method is employed:
\begin{equation}
\label{eq:Tve_incremental_midpoint}
\begin{array}{l}
\displaystyle
  - \int\limits_{\indLT{n}z}^{\indLT{n}z + \Delta z}
                2 \mu_k\,{\rm e}^{-\frac{\indLT{n}z + \Delta z-s}{\tau_k}} 
                \bigg(\dfrac{\rm d}{{\rm d}s}\Cg^\inv(s)\bigg) \ {\rm d}s
                \\[4mm]
\displaystyle
 \qquad\quad \approx -\Delta z \
                2 \mu_k\,{\rm e}^{-\frac{ \Delta z}{2\,\tau_k}} 
                \ \Cg!^\triangle^\inv\Big|_{\indLT{n}z + \frac{\Delta z}{2}} \ .
\end{array}               
\end{equation}
Furthermore, the material time derivative $\Cg!^\SLdreieck^\inv$ is approximated by
\begin{equation}
\label{eq:Tve_CG_approx}
\begin{array}{l}
  \Cg!^\triangle^\inv\Big|_{\indLT{n}z + \frac{\Delta z}{2}}
  \approx
    \dfrac{1}{\Delta z} \Big(\indLT{n+1}\Cg^\inv  - \indLT{n}\Cg^\inv \Big) \ .
\end{array}   
\end{equation}
A substitution of Eqs.~\eqref{eq:Tve_incremental_first_term} - \eqref{eq:Tve_CG_approx} into Eq.~\eqref{eq:Tve_incremental_split} yields the incremental representation of the stresses $\indLT{n+1}\Ttil_{ve,k}$ of one single Maxwell element
\begin{equation}
\label{eq:Tve_incremental_single}
\begin{array}{l}
  \indLT{n+1}\Ttil_{ve,k} 
      =  {\rm e}^{-\frac{\Delta z}{\tau_k}} \  \indLT{n}\Ttil_{ve,k} \, +
                \\[5mm]\displaystyle
    \qquad \qquad   \quad    
          - 2 \mu_k\,{\rm e}^{-\frac{ \Delta z}{2\,\tau_k}} 
                \ \Big(\indLT{n+1}\Cg^\inv  - \indLT{n}\Cg^\inv \Big) \ .
\end{array}
\end{equation}
Based on this result, the total viscoelastic stress contribution can be calculated by the help of Eq.~\eqref{eq:stress_visc}.

%
\subsection{ANSYS specific stress and material tangent}
\label{sec:FEM_implementation_ANSYS} 

The material model has been implemented into the commercial finite element software \textit{ANSYS}$^{\rm TM}$ by the help of the user subroutine \textit{USERMAT} \cite{Ansys_allg_2011}. Within a calculation step $\Delta t = t_{n+1} - t_{n}$, several input variables are transferred to the user subroutine. These are, for example, the deformation gradients $\indLT{n}\F$ and $\indLT{n+1}\F$ as well as user defined internal variables of the previous time step. In the opposite direction, appropriate output values have to be transferred to the software. The output includes the stress and the material tangent operator. Moreover, internal variables have to be updated for the next calculation step. 

\begin{figure}[ht]
	\centering
			\includegraphics[width=0.35\textwidth]{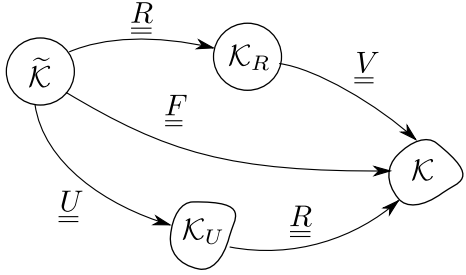}
	\caption{Polar decomposition of the deformation gradient.}
	\label{fig:defgrad_polar}
\end{figure}

The material model described in Sections \ref{sec:Curing_model} and \ref{sec:Curing_model_constitutive_functions}  has been set up in total Lagrangian representation. Thus, the stresses are given in the form of the $2^{\rm nd}$ Piola-Kirchhoff stress tensor $\Ttil$. Its corresponding material tangent operator $\STAPEL K!_\SLstrich!_\SLstrich!_\SLstrich!_\SLstrich!^\SLtilde$ is defined by
\begin{equation}
\label{eq:total_lagrange}
  \Ttil!^\SLdreieck = \STAPEL K!_\SLstrich!_\SLstrich!_\SLstrich!_\SLstrich!^\SLtilde \ppkt  \C!^\SLdreieck \ ,
  \qquad
  \STAPEL K!_\SLstrich!_\SLstrich!_\SLstrich!_\SLstrich!^\SLtilde 
     = \dfrac{\partial \Ttil}{\partial \C} \ .
\end{equation}
Since \textit{ANSYS}$^{\rm TM}$ uses an updated Lagrangian representation for the formulation of finite strain material models \cite{Ansys_allg_2011}, appropriate transformations of the stress and material tangent measures have to be performed. An  \textit{ANSYS}$^{\rm TM}$ specific stress tensor $\Ten2 \sigma_U$ and its corresponding material tangent operator $\STAPEL k!_\SLstrich!_\SLstrich!_\SLstrich!_\SLstrich_U$ are defined on the configuration $\mathcal{K}_U$ which arises from the polar decomposition theorem. (cf. Fig.~\ref{fig:defgrad_polar}). 
According to the polar decomposition theorem, the deformation gradient $\F$ gets decomposed into pure stretch tensors and a pure rotation tensor 
\begin{equation}
\label{eq:polar_decompo}
  \F = \Ten2 V \cdot \Ten2 R = \Ten2 R \cdot \Ten2 U \ .
\end{equation}
Therein, $\Ten2 V$ and $\Ten2 U$ are the positive definite symmetric left and right stretch tensor, respectively. Furthermore, $\Ten2 R$ is the orthogonal rotation tensor, thus $\Ten2 R^\inv = \Ten2 R^T$ holds. By the help of the stretch tensor $\Ten2 U$, the stress tensor $\Ten2 \sigma_U$ is obtained by the push forward operation of the  $2^{\rm nd}$ Piola-Kirchhoff stress tensor $\Ttil$ 
\begin{equation}
\label{eq:Cauchy_pull_back_final}
   \Ten2 \sigma_U 
     \,=\,  \dfrac{1}{J} \, \Ten2 U \cdot \Ttil \cdot \Ten2 U  
     \,=\,  \dfrac{1}{J} \, \STAPEL M!_\SLstrich!_\SLstrich!_\SLstrich!_\SLstrich_U  \ppkt \Ttil \ .
\end{equation}
Here, Eq.~\eqref{eq:S24} has been applied for the definition of the fourth order tensor
\begin{equation}
\label{eq:TensM}
   \STAPEL M!_\SLstrich!_\SLstrich!_\SLstrich!_\SLstrich_U
       =    \left( \Ten2 U \otimes \Ten2 U\right)^{S_{24}} \ .
\end{equation}
According to \cite{Ihlemann_2006} and \cite{Rendek_Lion_2010}, the corresponding material tangent $\STAPEL k!_\SLstrich!_\SLstrich!_\SLstrich!_\SLstrich_U$ is calculated by the operation
\begin{equation}
\label{eq:tangent_euler_rot_final}
   \STAPEL k!_\SLstrich!_\SLstrich!_\SLstrich!_\SLstrich_U
       =    \dfrac{2}{J} \ \STAPEL M!_\SLstrich!_\SLstrich!_\SLstrich!_\SLstrich_U  \ppkt
            \left\{
                \STAPEL K!_\SLstrich!_\SLstrich!_\SLstrich!_\SLstrich!^\SLtilde 
             + \left( \Ttil \otimes \C^\inv\right)^{S_{24}}
            \right\}
            \ppkt   \STAPEL M!_\SLstrich!_\SLstrich!_\SLstrich!_\SLstrich_U \ .
\end{equation}

Eqs. \eqref{eq:Cauchy_pull_back_final} and \eqref{eq:tangent_euler_rot_final} have been implemented into the user subroutine \textit{USERMAT} right after the computation of the Lagrangian tensors $\Ttil$ and $\STAPEL K!_\SLstrich!_\SLstrich!_\SLstrich!_\SLstrich!^\SLtilde $.

%
\subsection{Algorithmic correction of the initial volume}
\label{sec:FEM_implementation_correction}

In this section, an algorithm is presented which enables the consideration of initial conditions that differ from reference values used in constitutive equations for the representation of thermochemical volume changes. 

In Section \ref{sec:Constitutive_functions_Volume}, a material function $\varphi_{\theta C}(\theta,q)$ has been introduced which describes changes in volume related to heat expansion and chemical shrinkage processes. This function has been formulated with respect to a specific reference temperature $\thetatil$ and a reference value \mbox{$\curetil = 0$} for the degree of cure. If the current temperature $\theta(t)$ and the degree of cure $q(t)$ equal those reference values, the function yields \mbox{$\varphi_{\theta C}(\thetatil,\curetil) = 1$}. Thus, no change in density and volume is computed. If, however, the temperature or the degree of cure differ from their reference values, a certain volume change would be calculated depending on the differences of the input variables and the specific material properties (i.e. heat expansion and chemical shrinkage coefficients). The same situation may occur right at the beginning of a numerical simulation, i.e. in the initial state. In particular, the initial temperature  $\theta_0$ and the initial degree of cure $q_0$ may vary compared to the previously defined reference values $\thetatil$ and $\curetil$, respectively. In such a case, a value $\varphi_{\theta C}(\theta_0 \ne \thetatil,q_0\ne \curetil) \ne 1$ and, consequently, an immediate volume change would be calculated right at the beginning of a simulation. In finite element simulations, this would either lead the finite element mesh to change its volume or volumetric stresses would occur initially. Moreover, a distortion of the finite element mesh might occur.

\begin{figure}[ht]
	\centering
			\includegraphics[width=0.35\textwidth]{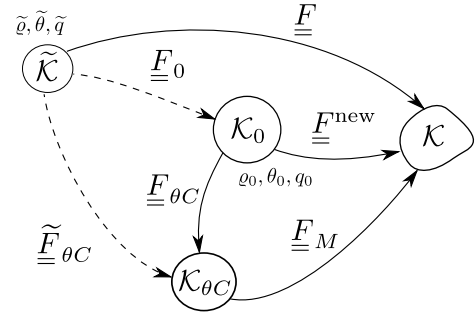}
	\caption{Decomposition of the deformation gradient for the distinction between reference and initial configuration}
	\label{fig:defgrad_korr}
\end{figure}

In order to avoid this undesired behaviour and to keep the initial volume of a finite element mesh constant, a correction is made at the beginning of a simulation. To this end, the reference state and the initial state at $t=0$ of a simulation are strictly separated by introducing a reference configuration $\Ktil$ and an initial configuration $\mathcal{K}_0$ according to Fig.~\ref{fig:defgrad_korr}.\footnote{The initial configuration $\mathcal{K}_0$ can be interpreted as a new reference \cite{Shutov_Etal_2012}.}

The reference configuration $\Ktil$ is constituted by reference values for the temperature $\thetatil$, the degree of cure \mbox{$\curetil = 0$} and the mass density $\rhotil$. Likewise, the initial state $\mathcal{K}_0$ at time $t=0$ of a simulation is represented by the initial values 
\begin{equation}
\theta_0 = \theta(t=0), \ q_0 = q(t=0), \ \varrho_0 = \varrho(t=0) \ . 
\end{equation}

Next, the different deformation paths occurring in Fig. \ref{fig:defgrad_korr} are defined. Firstly, a new deformation gradient $\F^{\rm new}$ is introduced  which represents the true deformation within a simulation. Accordingly, the deformation gradient $\F_{\theta C}$ represents the true thermochemical volume change. The latter operator is constituted by
\begin{equation}
\label{eq:F_thetaC_init}
  \F_{\theta C} = J_{\theta C}^{\,1/3} \ \I ,
\quad
  J_{\theta C} = \dfrac{{\rm d}V_{\theta C}}{{\rm d} V_0} = \dfrac{\varrho_0}{\varrho_{\theta C}} \ ,
\end{equation}
where $J_{\theta C}$ is the true thermochemical volume ratio that occurs in a simulation. Since the initial state $\mathcal{K}_0$ is assumed to be deformation free, the condition 
\begin{equation}
\label{eq:cond_init_vol}
\F^{\rm new}(t=0) \, = \, \F_{\theta C}(t=0) \, = \, \I 
\end{equation}
holds. Next, the mapping between the reference configuration $\Ktil$ and the initial configuration $\mathcal{K}_0$ is represented by an isotropic deformation gradient $\F_0$
\begin{equation}
\label{eq:F_0}
  \F_{0} 
     = J_0^{\,1/3} \ \I , \qquad
  J_0  
     = \dfrac{{\rm d}V_0}{{\rm d}\Vtil} 
     = \dfrac{\rhotil}{\varrho_0} \ .
\end{equation}
Here, $J_0$ represents the volume ratio between the reference and the initial configuration. This value is not a function of time and thus remains constant throughout the whole deformation process. Moreover, $J_0$ will be interpreted as a correction of the initial volume. 

Finally, a mapping between the reference state $\Ktil$ and the configuration $\mathcal{K}_{\theta C}$ is introduced as follows:
\begin{equation}
\label{eq:F_thetaC_ref}
  \F!^\SLtilde_{\theta C} = \varphi_{\theta C}^{1/3}\big(\theta,q\big) \ \I , \quad
  \varphi_{\theta C}\big(\theta,q\big) = \dfrac{{\rm d}V_{\theta C}}{{\rm d}\Vtil} = \dfrac{\rhotil}{\varrho_{\theta C}} \ .
\end{equation}
Note that $\F!^\SLtilde_{\theta C}$ is defined by the constitutive function  $\varphi_{\theta C}\big(\theta,q\big)$ (see Eq.~\eqref{eq:phi_thetaC}) and thus only depends on the current temperature $\theta(t)$ and degree of cure $q(t)$. It represents a hypothetical volume ratio that would occur if no correction is computed.

Next, the initial correction is calculated. According to Fig. \ref{fig:defgrad_korr}, the relation between the deformation gradients \eqref{eq:F_thetaC_init}, \eqref{eq:F_0} and \eqref{eq:F_thetaC_ref} at arbitrary values $\theta$ and $q$ reads as
\begin{equation}
\label{eq:F0_decompo}
  \F_{\theta C}(\theta, q) =  \F_0^\inv \cdot \F!^\SLtilde_{\theta C}(\theta, q) \ ,
\end{equation}
Since $\F_{\theta C}$, $\F_0$ and $\F!^\SLtilde_{\theta C}$ are isotropic, a similar relation can be formulated for the corresponding volume ratios:
\begin{equation}
\label{eq:J0_decompo}
 J_{\theta C}(\theta,q) =  \dfrac{\varphi_{\theta C}\big(\theta,q\big)}{J_0} \ .
\end{equation}
Recall that the function $\varphi_{\theta C}\big(\theta,q\big)$ only depends on the current temperature and the current degree of cure, and thus is fully determined. In contrast, the values $J_{\theta C}\big(\theta,q\big)$ and $J_0$ remain to be calculated. To this end, the condition \eqref{eq:cond_init_vol} is employed. It takes into account that no volume change occurs in the initial state. Thus, it is stated that the initial volume remains unaffected by different initial values $\theta_0$ and $q_0$. More precisely, 
\begin{equation}
\label{eq:init_vol}
  \F_{\theta C}(\theta_0, q_0) = \I 
  \ \ \Leftrightarrow  \ \ 
  J_{\theta C} (\theta_0, q_0) = 1 \ 
\end{equation}
is assumed. Based on this condition, the initial correction $J_0$ is calculated by evaluation of Eq.~\eqref{eq:J0_decompo} at the initial state $\theta = \theta_0$ and $q = q_0$, which yields
\begin{equation}
\label{eq:J0_result}
  J_{0} = \varphi_{\theta C}(\theta_0, q_0) = \text{const.} \ .
\end{equation}
Furthermore, the initial mass density $\varrho_0$ for a certain set of values $\theta_0$ and $q_0$ is adjusted by 
\begin{equation}
\label{eq:rho0_result}
  \varrho_{0} = \dfrac{\rhotil}{J_0}= \text{const.} \ .
\end{equation}
In summary, the algorithm works as follows. Firstly, the initial correction $J_0$ is calculated within the first load step by Eq.~\eqref{eq:J0_result}. This value is stored and can be accessed throughout all subsequent load steps of the simulation. Within each load step,  the function $\varphi_{\theta C}\big(\theta,q\big)$ constituted by Eq.~\eqref{eq:phi_thetaC} is evaluated and the true thermochemical volume ratio $J_{\theta C}\big(\theta,q\big)$ is computed by Eq. \eqref{eq:J0_decompo}.

%
\section{Finite element simulation of PMCs}
\label{sec:Finite_element_simulation}

In this section, the material model is applied within finite element simulations regarding the newly proposed manufacturing process for deep drawn PMCs (cf. Section~\ref{sec:Introduction}). Since this paper primarily focuses on phenomena related to the adhesive's curing reaction, the simulation of the forming step is reduced to a simplifying approximation based on a more complex forming simulation presented by Neugebauer \etal \cite{Neugebauer_Etal_2013_WGP}. Nevertheless, the forming step cannot be omitted since the subsequent volume shrinkage of the adhesive leads to quite different results concerning secondary deformations of the PMC or evolving strains on the MFC. 

The considered finite element model is related to one specific deep drawing geometry (see Section~\ref{sec:Finite_element_model}). Based on this model, two different manufacturing processes are investigated. Firstly, simulations on the new manufacturing process (see Section~\ref{sec:Introduction}) are conducted. To this end, Section~\ref{sec:Part_I_Forming_step} deals with the simplified forming step where the adhesive is not yet cured. The subsequent curing of the adhesive is considered in Section \ref{sec:Part_II_Curing_process}. To compare the obtained results regarding the impact on the formed PMC and the MFC, an alternative manufacturing process is investigated as well. In Section~\ref{sec:reversed_process} the order of forming and curing is switched which will illustrate the negative influence on the MFC when forming takes place after the adhesive has been fully cured.

\subsection{Finite element model}
\label{sec:Finite_element_model}

\begin{figure}[ht]
	\centering
  \includegraphics[width=0.45\textwidth]{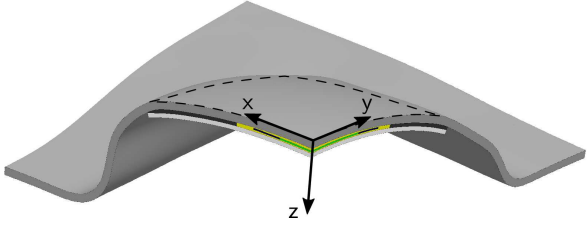}
	\caption{Quarter section of the deep drawn PMC with the section of the finite element model (dashed line) }
	\label{fig:pic_deep_drawing}
\end{figure}

The specific example of sheet metal forming simulation considered in this paper relies on a deep drawn rectangular cup geometry as presented in \cite{Neugebauer_Etal_2013_WGP}. A metal sheet with in-plane dimensions of $200 \, \rm mm$ x $130 \, \rm mm$ is to be formed. A cover metal sheet and a MFC are bonded to the structure by an adhesive as described in Section \ref{sec:Introduction}. A schematic illustration of a quarter of the final formed deep drawing cup is depicted in Fig.~\ref{fig:pic_deep_drawing}. 

\begin{figure}[ht]
	\centering
	\includegraphics[width=0.45\textwidth]{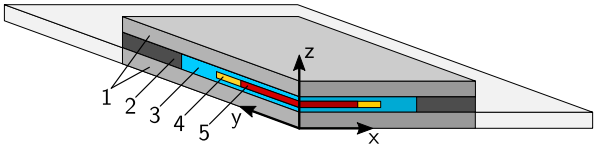}
	\caption{Geometry of the model with aluminium layers (1), 
					 spacer (2), adhesive layer (3) and 
					 macro fibre composite (4) with its active part (5)}
	\label{fig:pic_model}
\end{figure}

Deduced from the objectives of this work, the finite element model used in this work is confined to the MFC's surrounding region of the PMC (see the dashed line in Fig.~\ref{fig:pic_deep_drawing}). Furthermore, two planes of symmetry are utilized such that the final model covers a quarter of the inner part of the PMC. Its basic area coincides to the size of the quarter aluminium cover sheet (Fig.~\ref{fig:pic_model}). Taking into account the thicknesses of the different layers (see Table~\ref{tab:Thicknesses}),  the overall dimensions of the employed finite element model are $42.5 \, \rm mm$ x $35 \, \rm mm$ x $2.9 \, \rm mm$.

\begin{table}[ht]
   \caption{Thicknesses of the PMC layers in the finite element model}
    \vspace{.5ex}
  \centering
  {\begin{tabular}{p{.05cm}p{4.4cm}p{2cm}p{.05cm}}
    \hline & & & \\[-3mm]
    &layer & thickness&\\ 
    \hline & & & \\[-3mm]
      &aluminium cover sheet  & $0.8  \ \rm mm$ &\\[1mm]
      &adhesive above MFC     & $0.15 \ \rm mm$ &\\[1mm]
      &macro fibre composite  & $0.3  \ \rm mm$ &\\[1mm]
      &adhesive below MFC     & $0.15 \ \rm mm$ &\\[1mm]
      &aluminium bottom sheet & $1.5  \ \rm mm$ &\\[1mm]
    \hline
  \end{tabular}}
  \label{tab:Thicknesses}
\end{table} 

Both, the PMC and the finite element model consist of several components with remarkably different material behaviour. Therefore, different material models are applied for the layers of the PMC. To represent the curing behaviour of the adhesive layer, the model and its corresponding material parameters presented in Section \ref{sec:Curing_model} are employed. The material behaviour of the aluminium sheets is described by an elastic-plastic model provided by \textit{ANSYS}$^{\text{TM}}$ \cite{Ansys_allg_2011}. The model incorporates plastic hardening effects represented by a bilinear kinematic hardening rule.\footnote{ For even more exact prediction of the residual stresses and spring back, material models with nonlinear kinematic hardening are needed \cite{Shutov_Kreissig_2008}.} The corresponding material parameters are chosen according to the manufacturer's specifications \cite{ENAW5083_2002}. The active part of the MFC with its unidirectional piezoceramic fibres aligned in y-direction is represented by a model of transversely isotropic elasticity \cite{Giddings_2009,MFC_2012}. The non-active part of the MFC as well as the spacer are modelled by isotropic elasticity laws. A summary of the employed material parameters for the described materials is given in Table~\ref{tab:MatPar_FE_Model}.

\begin{table}[ht]
  \caption{Material parameters for the different layers of the PMC}
  \vspace{.5ex}
  \centering
  {\begin{tabular}{|lll|}
  \hline && \\[-3mm]
  \multicolumn{3}{|l|}
     {\underline{Metal sheets: Elastoplasticity with kinematic hardening}}\\[3mm]
      & $E \,\,\, = \ 70000 \ \rm MPa$ 
      & $\ \nu  \,\,\, = \ 0.31$                              \\[1mm] 
      & $\sigma_F = \      150 \ \rm MPa$      
      & $\  E_T  = \ 1015 \ \rm \rm MPa$     \\[4mm]
  \multicolumn{3}{|l|}
     {\underline{Active part of the MFC: Transversely isotropic elasticity} }\\[3mm]
      & $E_y \, \; = \ 30336 \ \rm MPa$
      & $\ E_x  \, \; = \ E_z = 15857 \ \rm MPa$ \\ [1mm]
      & $\nu_{xz} \,\, = \  0.31$         
      & $\ \nu_{xy} \,\, =\  \nu_{zy}= 0.16$          \\[1mm]
      & $G_{xy} = \ G_{zy}= 5515 \ \rm MPa$    
      & $\ G_{zx} = \ \frac{E_{x}}{2\left( 1 + \nu_{xz} \right)}$       \\[4mm] 
  \multicolumn{3}{|l|}
     {\underline{Non-active part of the MFC: Isotropic elasticity}}\\[3mm]
      & $E_{kapt} = \ 3500 \ \rm MPa$ 
      & $\  \nu_{kapt}  = \ 0.33$         \\[4mm]
  \multicolumn{3}{|l|}
     {\underline{Spacer: Isotropic elasticity}}\\[3mm]
      & $E_{tape} = \ 8000 \ \rm MPa$ 
      & $\  \nu_{tape}  = \ 0.35$         \\[1mm]
   \hline
  \end{tabular}}
  \label{tab:MatPar_FE_Model}
\end{table}

\begin{figure}[ht]
	\centering
	\includegraphics[width=0.45\textwidth]{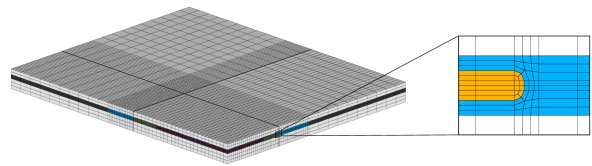}
	\caption{Finite element mesh with highlighted radii at the outer edge of the MFC}
	\label{fig:pic_Mesh}
\end{figure}

The finite element model has been meshed by a bottom up approach with three-dimensional structural solid elements, each incorporating eight nodes and linear regression functions (Fig.~\ref{fig:pic_Mesh}). The complete mesh consists of about $40000$ elements from which $14000$ involve the adhesive's material model. To avoid volume locking effects within the adhesive layer,  a mixed u-p-formulation is used for the corresponding elements. Furthermore, radii at the outer edges of the MFC have been modelled to reduce effects of stress concentration (see highlighted region of Fig.~\ref{fig:pic_Mesh}).

\subsection{Part I: Forming step}
\label{sec:Part_I_Forming_step}

\begin{figure}[ht]
	\centering
	\includegraphics[width=0.3\textwidth]{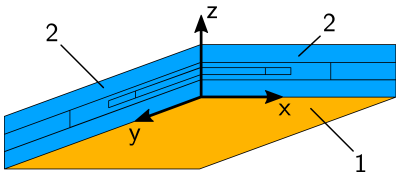}
	\caption{Interpolated displacement values at the bottom (1) 
	         and symmetry boundary conditions (2) on the model 
	         for the simulation of the forming step}
	\label{fig:pic_boundary_conditions_I}
\end{figure}

The first step of the simulation process approximates the deep drawing of the PMC's inner part. To this end, node displacements from the deep drawing simulation in \cite{Neugebauer_Etal_2013_WGP} have been interpolated and defined as boundary conditions on the bottom sheet (see Fig.~\ref{fig:pic_boundary_conditions_I}). This approach of approximating the deformation of the inner part provides the possibility to focus on studies regarding regions surrounding the adhesive while reducing the computational effort of a complete deep drawing simulation. Thus, typical challenges in sheet metal forming simulations like wrinkling, spring-back and contact formulations are avoided.

Within the forming step, the adhesive has not yet reached its gelation point and, thus, can be treated as a viscoelastic fluid \cite{Winter_1986}. To adopt the fluid like material behaviour, a simplified model is used which includes pure elastic behaviour with respect to volume changes and viscoelastic behaviour in case of isochoric deformations. The viscoelastic behaviour is represented by one single Maxwell element with constant material parameters. Here, an \textit{ANSYS}$^{\rm TM}$ built in viscoelastic model has been used. The material parameters are $K = 5000 \ \rm MPa$ for the bulk modulus, $c = 2.0 \ \rm MPa$ for the Neo-Hookean stiffness, and $\tau = 20.0 \ \rm s$ for the relaxation time. 

The simulation time of the forming step is $1000 \ \rm s$. The forming itself takes $30 \ \rm s$. The remaining period of $970 \ \rm s$ is included to achieve a relaxed state in the adhesive. This procedure has been chosen due to numerical difficulties when using shorter simulation times for the forming step or applying smaller values for the adhesive's relaxation time. Some numerical investigations on appropriate representations of the liquid adhesive during forming were conducted in \cite{Neugebauer_Etal_2013_WGP}. However, the aim of this simulation step is to obtain a finite element mesh of the formed PMC. Thus, the described procedure is assumed to be sufficient for the needs of this work. 

\begin{figure}[ht]
	\centering
	\includegraphics[width=0.45\textwidth]{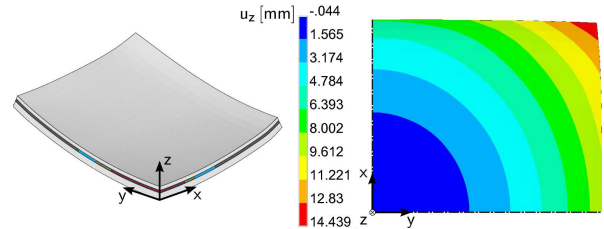}
	\caption{Perspective view (left) and bottom view (right) of the 
					 model's $u_z$-displacement due to forming}
	\label{fig:pic_uz_bottom}
\end{figure}

Resulting from the forming simulation, Fig.~\ref{fig:pic_uz_bottom} shows the circularity of the contour lines of the displacement $u_z$ which points to a good reproduction of the profile generated by the rectangular punch with a double curvature of $100 \, \rm mm$.

In order to analyse the functionality of the formed PMC, the strain affecting the MFC has to be examined. According to manufacturer's specifications, the linear elastic tensile strain limit of the MFC's brittle piezoceramic fibres is about  $1 \! \cdot \! 10^{-3}$~\cite{Daue_Kunzmann_2010,MFC_2012}. If this strain level is exceeded, risk of depolarization effects increases significantly and the MFC might possibly not be able to maintain its sensor and actuator functionalities in the required magnitude ~\cite{Daue_Kunzmann_2010}. Complete failure of the MFC occurs, if a maximum operational tensile strain of $4.5 \! \cdot \! 10^{-3}$ is exceeded \cite{MFC_2012}. 

\begin{figure}[ht]
	\centering
	\includegraphics[width=0.45\textwidth]{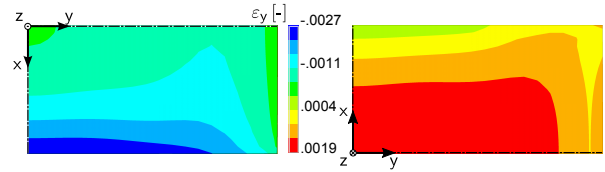}
	\caption{Total mechanical strain $\varepsilon_{y}$ along the orientation of fibres on the top (left) and the bottom (right) of the active part of the MFC}
	\label{fig:pic_MFC_eps_y}
\end{figure}

As a result of the conducted forming simulation,  Fig.~\ref{fig:pic_MFC_eps_y} reveals that the deformation of the forming step exceeds the linear tensile strain limit by two to three times. However, failure of the MFC is not predicted since the strain magnitudes are below the maximum operational tensile strain. Moreover, it can be seen that there are compressed and stretched regions which points to major influence of bending deformation on the MFC. 

\subsection{Part II: Curing process}
\label{sec:Part_II_Curing_process}

The second part of the simulation gives insight into the impact of the adhesive's curing and its associated volume shrinkage on the PMC. In order to obtain a continuous process chain of simulation, the deformed mesh of the simplified forming simulation presented in Section \ref{sec:Part_I_Forming_step} is employed as starting point for the curing simulation. To account for curing phenomena in the adhesive, the viscoelastic material model used within the forming step is replaced by the material model presented in Sections~\ref{sec:Curing_model} and \ref{sec:Curing_model_constitutive_functions}. The point of time to conduct this change of material laws is set to the gelation point at which the fluid turns into a solid \cite{Winter_1986}. According to Meuwissen \etal \cite{Meuwissen_Etal_2004} the gelation point is represented by a degree of cure \mbox{$q \approx 0.5-0.6$}. Here, the initial value is set to $q_0 = 0.5$. 

\begin{figure}[ht]
	\centering
	\includegraphics[width=0.45\textwidth]{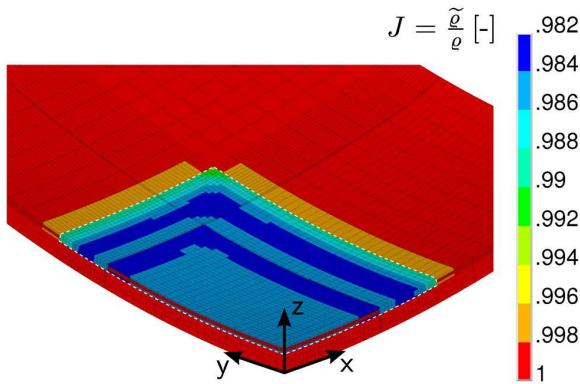}
	\caption{Volume ratio~$J$ of the PMC after the curing process shows the adhesive's chemically induced shrinkage (cover sheet is hidden, highlighted region indicates the adhesive)}
	\label{fig:pic_J3_section}
\end{figure}

Within the curing simulation, the applied boundary conditions are confined to symmetries at the two cross-sectional planes as can be seen in Fig.~\ref{fig:pic_boundary_conditions_I}.  The boundary conditions on the bottom sheet are released. Since no residual stresses are transferred from the forming simulation to the curing step, no initial spring-back is observed. The simulation time of the curing simulation is set to $1800 \, \rm s$ according to manufacturer's specifications \cite{DP410_2003}. Additionally, a constant temperature $\theta = 318 \, K$ is prescribed. As an example for the decisive effects of adhesive curing process, Fig.~\ref{fig:pic_J3_section} shows the mechanical volume ratio~$J$ resulting from the material's volume shrinkage. The final degree of cure at this stage is $q = 0.89$.

\begin{figure}[ht]
	\centering
	\includegraphics[width=0.45\textwidth]{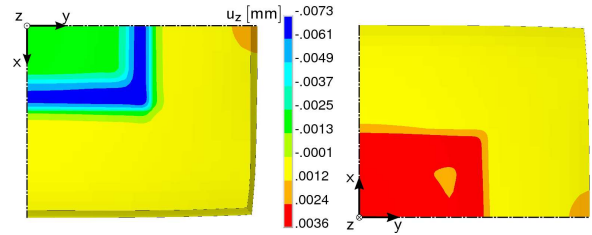}
	\caption{Surface markings due to the adhesive's volume shrinkage during the curing process on top (left) and at the bottom (right) of the finite element model}
	\label{fig:pic_uz_curing}
\end{figure}

Note that if a free volume shrinkage is assumed, $J$ would be equally distributed  throughout the entire adhesive volume. The apparent deviations of~$J$ shown in Fig.~\ref{fig:pic_J3_section} are caused by supporting effects of adjacent materials. These supporting effects can also be observed by viewing the displacements in normal direction of the free surfaces of the aluminium layers (see Fig.~\ref{fig:pic_uz_curing}). As a consequence of their different thicknesses, the resulting normal displacements at the free surface of the top layer are more pronounced than these of the bottom layer. 

\begin{figure}[ht]
	\centering
	\includegraphics[width=0.45\textwidth]{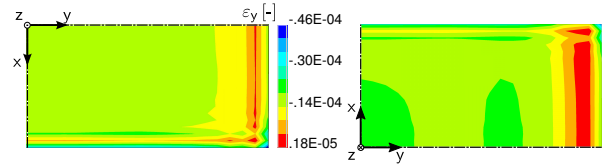}
	\caption{Total mechanical strain~$\varepsilon_{y}$ in the direction of the orientation of the fibres on the top (left) and the bottom (right) of the MFC due to the chemical curing process}
	\label{fig:pic_MFC_eps_y_curing}
\end{figure}

Analogous to the investigations of the simplified for\-ming simulation (cf. Section \ref{sec:Part_I_Forming_step}), the mechanical strain $\varepsilon_{y}$ of the piezoceramic fibres is examined and presented in Fig.~\ref{fig:pic_MFC_eps_y_curing}. 

As can be seen from Fig.~\ref{fig:pic_MFC_eps_y_curing}, the linear elastic tensile strain limit is not exceeded due to the curing reaction. However, superimposing the strains of the forming and the curing simulation reveals that the MFC is loaded additionally in the compressed zones. For the herein presented example with its adhesive layer thickness, its further geometrically dimensions and the given forming displacements, the influence is rather negligible. Nevertheless, deviations from these conditions (smaller radii or thicker adhesive layer) may result in strain amplitudes due to forming and curing process which are vice versa.

\subsection{Comparison to reversed manufacturing process}
\label{sec:reversed_process}

\begin{figure}[ht]
	\centering
	\includegraphics[width=0.45\textwidth]{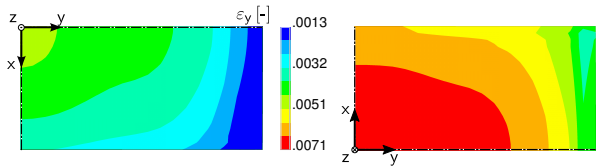}
	\caption{Total mechanical strain~$\varepsilon_{y}$ along fibre orientation on the top (left) and the bottom (right) of the MFC due to forming with already cured adhesive}
	\label{fig:pic_MFC_eps_y_reverse}
\end{figure}

Finally, an analysis of a reversed manufacturing process proves the adhesive's protecting function to the MFC. The concluding simulation is conducted with the material model presented in Sections~\ref{sec:Curing_model} and \ref{sec:Curing_model_constitutive_functions}. Here, an initial degree of cure of~$q_0=0.5$ is chosen and a constant temperature of $\theta = 318 \,K$ is prescribed. The simulation time of the curing process is $1800 \, \rm s$. Within this period of time, the chemical reaction is finished with the value  $q = 0.89$ which is in accordance with the simulation in Section \ref{sec:Part_II_Curing_process}. Subsequently, the forming step with the cured adhesive is simulated. In Fig.~\ref{fig:pic_MFC_eps_y_reverse}, the resulting tensile strain along the fibre direction is depicted.

As Fig.~\ref{fig:pic_MFC_eps_y_reverse} reveals, due to this strategy the entire MFC has been uniformly stretched and the maximum strain is almost four times higher than in the simulation of the actually intended manufacturing process shown in Sections~\ref{sec:Part_I_Forming_step} and \ref{sec:Part_II_Curing_process}. Moreover, not only the linear elastic tensile strain limit but also the maximum operational tensile strain of $4.5 \! \cdot \! 10^{-3}$ is exceeded. Hence, failure of the MCF is predicted in this simulation. In conclusion, it can be confirmed that the liquid adhesive protects the MFC during the forming process (see results in Sections \ref{sec:Part_I_Forming_step} and \ref{sec:Part_II_Curing_process}).
%
\section{Conclusion and discussion}
\label{sec:Conclusions}

The present work aims at supporting investigations of the innovative manufacturing process for smart PMCs. Apparently, the role of the adhesive is crucial due to its specific material behaviour and its geometrical design. To allow for investigations on this issue, a general material modelling approach is presented in Section \ref{sec:Curing_model}. Furthermore, a concretized model for the representation of curing phenomena in one specific adhesive is described in Section \ref{sec:Curing_model_constitutive_functions}. This model is able to capture the main characteristics of the two-component epoxy based adhesive considered in this paper. Moreover, the thermodynamic consistency of the model could be proved. 

In Section \ref{sec:FEM_implementation}, different aspects of numerical implementation into the finite element software \textit{ANSYS}$^{\rm TM}$ have been discussed. Beside the numerical integration of constitutive equations and the derivation of software specific stress and material tangent tensors, a new algorithm has been proposed which suppresses  undesired volume changes at the beginning of numerical simulations. Those volume changes may result from thermal expansion and chemical shrinkage when initial values for the temperature and the degree of cure differ from their reference values. Only by the help of the new algorithm, the consideration of heterogeneous initial fields of temperature and degree of cure has been made possible.

Finally, in Section \ref{sec:Finite_element_simulation} a finite element model for a deep drawn rectangular cup has been built up as representative example for the innovative manufacturing process. The model is based on experimental and numerical studies presented by Neugebauer \etal \cite{Neugebauer_Etal_2013_WGP}. However, in the present work a simplified model which includes only the closest surroundings to the MFC has been employed. Based on the results of the forming and curing simulations, it can be stated that the simplified forming step is capable of producing a sufficiently accurate geometry as basis for the curing simulation. 

To highlight the benefit of the new manufacturing process, a reversed sequence of production has been investigated as well and the results of both processes have been compared. In Section \ref{sec:reversed_process} it has been predicted that failure of the MFC occurs if the PMC is formed after the adhesive is completely cured. Hence, it can be seen that the new manufacturing process considered in this paper exhibits particular potential to overcome these shortcomings. Here, the floating support resulting from the uncured adhesive prevents sufficiently from overloading or delamination during the forming step.

Based on the present work, which shows the qualitative feasibility of the proposed material model as well as the strategy of numerical simulation, different aspects are planned to be conducted in the future. This includes  complete identification of material parameters based on experimental investigations and the extension to thermomechanically coupled simulations analogous to \cite{Landgraf_Etal_2012,Landgraf_Etal_2013} or \cite{Mahnken_2013}. Finally, it is intended to employ the presented approach to the simulation of different deep drawing processes (cf. \cite{Neugebauer_Etal_2010_ProdEng,Neugebauer_Etal_2013_WGP}).


\begin{acknowledgements}
This  research  is  supported  by  the  German Research Foundation 
 (DFG) within the Collaborative  Research  Centre/Transregio  39 
PT-PIESA, and the collaborative research project DFG PAK 273. 
This support is greatly acknowledged.
\end{acknowledgements}

\bibliographystyle{spmpsci}      
\bibliography{lit_Curing_PMC_Landgraf_Etal}   


\appendix
%
\section{Remark on thermodynamic consistency}
\label{sec:Constitutive_functions_Process_consistency}

In Section \ref{sec:Curing_model_general_framework} the thermodynamic consistency of the general modelling framework has been considered but not finally proved since specific constitutive functions had not been set up to that point. The evaluation of the two remaining conditions \eqref{eq:dissi_qdot} and \eqref{eq:dissi_zdot} is discussed in this section. To this end, constitutive assumptions presented in Sections \ref{sec:Constitutive_functions_Degree_of_Cure} - \ref{sec:Constitutive_functions_Process_dependency} are employed.

Firstly, inequality~\eqref{eq:dissi_zdot} is considered. To prove this condition, the partial derivative of the isochoric part of the free energy function $\hat{\psi}_G$ with respect to the intrinsic time scale $z$ has to be calculated. Since only the viscoelastic parts $\hat{\psi}_{ve,k}$ include the dependency on $z$ (cf. Eq.~\eqref{eq:psi_G}), it remains to show that
\begin{equation}
\label{eq:therm_cons_psi_ve}
   - \sum_{k=1}^{N_k} \dfrac{\partial \hat{\psi}_{ve,k}}{\partial z} \ge 0 \ .
\end{equation}
Furthermore, it is assumed that not only the sum of all Maxwell elements but also every single Maxwell element meets the consistency condition. Thus, it is sufficient to show that
\begin{equation}
\label{eq:therm_cons_psi_ve_k}
  - \dfrac{\partial \hat{\psi}_{ve,k}}{\partial z} \ge 0
\end{equation}
holds. According to \cite{Haupt_Lion_2002} or \cite{Lion_Kardelky_2004}, the thermodynamic consistency of the ansatz \eqref{eq:psi_visc_single} is met, if the conditions 
\begin{equation}
\label{eq:therm_cons_psi_ve_conditions}
   G_k(t)  \ge 0 \ ,
  \qquad
  \dfrac{\rm d}{{\rm d}t} G_k(t)  \le 0 \ ,
  \qquad
  \dfrac{\rm d^2}{{\rm d}t^2} G_k(t)  \ge 0
\end{equation}
hold. Obviously, these conditions are satisfied by the relaxation function \eqref{eq:kernel}.

Next, the remaining condition \eqref{eq:dissi_qdot} is considered. Since inequality \eqref{eq:dissi_qdot} cannot be evaluated in a general form, an estimation under consideration of some physically reasonable assumptions is employed. In  the first step, the term $\partial \hat{\psi}_{\theta C}(\theta,q) / \partial q $ of Eq.~\eqref{eq:dissi_qdot} is estimated. Here, an ansatz for the thermochemical part of the specific enthalpy per unit mass is introduced \cite{Kolmeder_Etal_2011,Lion_Yagimli_2008}
\begin{equation}
\label{eq:ansatz_enthalpy}
  h_{\theta C}(\theta,q) = h_{fluid}(\theta) \, (1-q) + h_{solid}(\theta) \, q \ .
\end{equation}
The functions $h_{fluid}(\theta)$ and $h_{solid}(\theta)$ are the specific enthalpy per unit mass of the uncured and fully cured material, respectively. Note that the general  ansatz \eqref{eq:ansatz_enthalpy} depends on the temperature $\theta$ and the degree of cure $q$.  Specific models for the consideration of temperature dependent behaviour have been introduced in \cite{Kolmeder_Etal_2011} and \cite{Lion_Yagimli_2008}. However, for the estimation conducted in this section this is omitted and the values for $h_{fluid}$ and $h_{solid}$ are assumed to be constant.  

The next step is to calculate the thermochemical free energy $\psi_{\theta C}$ from $h_{\theta C}$. This can be accomplished by approaches presented in \cite{Lion_Yagimli_2008} or \cite{Mahnken_2013}. However, here an alternative formulation of this calculation step is used as follows. Firstly, the Legendre transformation 
\begin{equation}
\label{eq:legendre}
  \psi + \theta \, \eta = h + \dfrac{1}{\rhotil} \ I_1(\Ttil\cdot \Ten2 \gamma) 
\end{equation}
is employed which relates the free energy and the enthalpy (see, for example, \cite{Lion_Yagimli_2008,Lubarda_2004}). Therein, $\eta$ is the specific entropy per unit mass and $\Ten2 \gamma = (\C-\I)/2$ is the Green-Lagrange strain tensor. Next it is assumed, that DSC experiments take place at zero mechanical stresses \cite{Lion_Yagimli_2008}. Thus, the last term of \eqref{eq:legendre} is neglected. Furthermore, the constitutive relation $\eta = - \partial \psi / \partial \theta$ at constant stress state is employed and the resulting equation is formulated with respect to the thermochemical potentials $\psi_{\theta C}$ and $h_{\theta C}$. This yields the reduced relation 
\begin{equation}
\label{eq:potentials_dgl}
  \psi_{\theta C}(\theta,q) - \theta \, \dfrac{\partial \psi_{\theta C}(\theta,q)}{\partial \theta} = h_{\theta C}(\theta,q) \ .
\end{equation}
Eq.~\eqref{eq:potentials_dgl} is a differential equation which has to be solved for $\psi_{\theta C}$. Its solution reads as
\begin{equation}
\label{eq:potentials_dgl_solu}
   \psi_{\theta C}(\theta,q) = C \, \dfrac{\theta}{\theta_0} - \theta \, \int \dfrac{1}{\theta^2}\,h_{\theta C}(\theta,q) \, {\rm d}\theta \ .
\end{equation}
Here, $C$ is an integration constant that does not need to be determined in our evaluation. Next, this general solution is applied to the ansatz for the thermochemical enthalpy which has been introduced in Eq.~\eqref{eq:ansatz_enthalpy}. This yields a specific model for the thermochemical free energy 
\begin{equation}
\label{eq:potentials_dgl_solu_specific}
   \psi_{\theta C}(\theta,q) = C \, \dfrac{\theta}{\theta_0} + h_{fluid}\, (1-q) + h_{solid} \, q \ .
\end{equation}
Based on this solution, the term $\partial \hat{\psi}_{\theta C}(\theta,q) / \partial q $ of Eq. \eqref{eq:dissi_qdot} is calculated by
\begin{equation}
\label{eq:dpsi_dq_solu}
   \dfrac{\partial \hat{\psi}_{\theta C}(\theta,q)}{\partial q} 
      =  \rhotil  \ \dfrac{\partial \psi_{\theta C}(\theta,q)}{\partial q} 
      = \rhotil  \ (h_{solid} - h_{fluid}) \ .
\end{equation}
To quantify this expression, the maximum specific reaction enthalpy per unit mass $\Delta h$ of a complete curing experiment has to be taken into account. This quantity has been measured by DSC experiments (see \cite{Kolmeder_Etal_2011,Lion_Yagimli_2008} for detailed description) and can be related to the model \eqref{eq:ansatz_enthalpy} by the relation
\begin{equation}
\label{eq:hsolid_hfluid}
   \Delta h = h(\theta,q=1) - h(\theta,q=0) = h_{solid} - h_{fluid} \ .
\end{equation}
Here, a value of $\Delta h \approx - 300 \, \rm J/g$ has been identified. Furthermore, taking into account the mass density $\rhotil \approx 1.1 \, \rm g/cm^3$, the first term of Eq.~\eqref{eq:dissi_qdot} is estimated by 
\begin{equation}
\label{eq:dpsidq_value}
\partial \hat{\psi}_{\theta C}(\theta,q) / \partial q  = -330 \ \rm MPa \ .
\end{equation}

Next, the second term in inequality \eqref{eq:dissi_qdot} is examined. Therein, the chemical shrinkage parameter $\beta_q$ can be identified by the help of Eq.~\eqref{eq:phi_thetaC}. Here, the relation
\begin{equation}
\label{eq:calc_betaq}
 \dfrac{1}{J_{\theta C}}\,
           \dfrac{\partial J_{\theta C}}{\partial q} \,
  = \beta_q 
\end{equation}
holds. Furthermore, a relation to the hydrostatic pressure $p$ is obtained by evaluation of
\begin{equation}
\label{eq:calc_pressure}
   \dfrac{\partial \hat{\psi}_V}{\partial J_M} \, J_M= - J p\ ,
  \quad
  p = - \dfrac{1}{3} \, \dfrac{1}{J} \, I_1(\Ttil\cdot\C) \ .
\end{equation}

Finally, inequality \eqref{eq:dissi_qdot} can be evaluated. To this end, expressions \eqref{eq:calc_betaq} and \eqref{eq:calc_pressure} are substituted into Eq. \eqref{eq:dissi_qdot} which yields
\begin{equation}
\label{eq:reformulate_cond}
 - \dfrac{\partial \hat{\psi}_{\theta C}(\theta,q)}{\partial q}
       - \beta_q\,J\,p \ \ge 0 \ .
\end{equation}
Moreover, the estimation \eqref{eq:dpsidq_value}  and the chemical shrinkage parameter $\beta_q=-0.05$ (cf. Section \ref{sec:Constitutive_functions_Volume}) are inserted in \eqref{eq:reformulate_cond}, and the resulting inequality is resolved for the expression $J\,p$. This finally yields the condition
\begin{equation}
\label{eq:pressure_cond}
  J\,p  \ \ge - 6600 \ {\rm MPa} \ .
\end{equation}
Since $J>0$ holds in general, it can be concluded from \eqref{eq:pressure_cond} that a hydrostatic pressure with $p>0$ does not endanger the thermodynamic consistency. However, if the material is loaded in hydrostatic tension ($p<0$), the condition \eqref{eq:pressure_cond} may be violated. If a constant volume is assumed ($J=1$), a hydrostatic tension of $p = -6600 \, \rm MPa$ would be necessary to violate thermodynamic consistency. Nevertheless, this value seems to be unrealistic to be achieved in real experiments. Thus, the thermodynamic consistency can be proved for the case of physically reasonable conditions (see also \cite{Lion_Hoefer_2007,Lion_Yagimli_2008,Mahnken_2013}).

\end{document}